\documentclass{aa}

\usepackage{graphicx}
\usepackage{multirow}  
\usepackage{txfonts}
\usepackage{academicons}

\usepackage{natbib,twoopt}
\usepackage[dvipsnames]{xcolor}

\newcommand{\Gaia}{{\it Gaia}}

\newcommand{\gspspec}{{\it GSP-Spec}}

\newcommand{\T}{$T_{\rm eff}$}
\newcommand{\g}{log($g$)}
\newcommand{\gunits}{cm/s$^2$}
\newcommand{\meta}{[M/H]}
\newcommand{\alfa}{$\alpha$}
\newcommand{\AF}{[\alfa/Fe]}

\newcommand{\SNR}{$S/N$}

\newcommand{\EBpRp}{$E_{(B_P-R_P)}$}
\newcommand{\AG}{$A_G$}

\usepackage{listings}
\definecolor{dkgreen}{rgb}{0,0.6,0}
\definecolor{gray}{rgb}{0.5,0.5,0.5}
\definecolor{mauve}{rgb}{0.58,0,0.82}

\begin{document} 

   \title{The \Gaia\ \gspspec\  catalogue of interstellar extinctions, and stellar luminosities, radii, and masses}
   \titlerunning{The \gspspec\ catalogue of \EBpRp, \AG, L, R and M}

   \author{
          Patrick de Laverny \inst{1}
          \and     
          Alejandra Recio-Blanco\inst{1}
          \and           
          Camila Navarrete \inst{1}
          \and
          Pedro A. Palicio \inst{1}
          \and
          Emanuele Spitoni \inst{2}
          } 

   \institute{
   Université Côte d'Azur, Observatoire de la Côte d'Azur, CNRS, Laboratoire Lagrange, Bd de l'Observatoire, CS 34229, 06304 Nice cedex 4, France  
   \and
   I.N.A.F., Osservatorio Astronomico di Trieste, via G.B. Tiepolo 11, 34143, Trieste, Italy
   }

   \date{Received ??, 2026 ; accepted ??}
   
   \abstract 
   {The \Gaia/DR3 \gspspec\ module has published the atmospheric parameters of up to 5.6 million stars based on the analysis of their Radial Velocity Spectrometer spectra. By combining these spectroscopic parameters with \Gaia\ parallax and photometric measurements, a large catalogue of interstellar colour excesses and extinctions, complemented with stellar luminosities, radii, and masses, was constructed without relying on any stellar evolution or structure models. This catalogue also contains the associated uncertainties estimated from Monte-Carlo realisations for about 4.6 million stars. We present a system of quality flags based on the \gspspec\ quality parameters, the achieved numerical precision, and the \Gaia\ astrometric quality. Adopting these flags, we defined a subsample of high-accuracy and precision parameters of more than 1.5 million stars. The impact of possible \gspspec\ parameter inaccuracies on the derived extinctions, luminosities, radii, and masses was also explored and revealed that the mass is the most affected quantity.  The radii and masses were validated by comparison with interferometric and asteroseismic data. This confirmed their high quality, even when the targets were highly extincted by the interstellar medium. We also emphasise that they are fully compatible and homogeneous with the \gspspec\ parameters. This allowed us to avoid systematics and biases that might be hidden when data from different (and potentially) heterogeneous catalogues are combined. Finally, we present some example applications of this catalogue: (i) exoplanet radii and masses, (ii) present-day mass distribution functions,  (iii) identification of Galactic populations based on their mass distribution, and (iv) Galactic halo accreted stellar luminosities and masses that confirm their merger epochs.}
   
   \maketitle

\section{Introduction}
\label{Sec:Intro}

Stellar parameters are essential for many fields in astrophysics, from exoplanetary, stellar, and Galactic to extragalactic studies. In general, the main stellar atmospheric parameters
(effective temperature \T, surface gravity \g, and mean metallicity \meta), plus the individual chemical abundances are estimated from the analysis of the light received from stars with photometric and/or spectroscopic methods. However, other fundamental 
stellar parameters cannot be directly derived from observations. This is generally the case, in particular, for the stellar radii ($R$) and masses ($M$), not to mention ages. 
Atmospheric parameters and chemical abundances depend on our knowledge of the physics of the studied stellar object. However, other fundamental parameters additionally depend on our capacity to predict how these physics evolve with time. For example, these quantities are often derived by adopting specific model-dependent technics, such as isochrone fitting for the mass and ages \citep[see, for instance,][]{Twins}.
These projection techniques are most often (and better) applied to main-sequence turn-off or subgiant stars because their atmospheric parameters strongly depend on time for these specific evolutionary
stages. Another successful method for deriving these stellar fundamental parameters is asteroseismology based on space observations, which can
provide  much better accuracies, but for only a few thousand red giants \citep[among others]{APOK2, APOKASC3}.  The analysis of
the stellar oscillations can indeed provide an accurate radius and mass,
through seismic scaling relations and knowing \T\ \citep[see, for instance,][]{Miglio13}. However, these asteroseismic data are not available for large samples of stars at any stage of their evolution, and in particular, for the main-sequence
stars \citep{APOKASCnaines}. For accurate stellar radius measurements, we might also cite interferometric techniques, which can provide good results, but for a rather small number of close
stars with well-known distances \citep{Salsi20}. 

All of these methods depend on the knowledge of the stellar atmospheric parameters, in particular, \T\ and \meta. In this context, the ESA \Gaia\ mission has revolutionised the availability of these stellar parameters 
by providing accurate and precise parallaxes for several hundred million stars \citep[see, for the third data release,][]{GaiaDR3}.
For a small fraction of them, whose spectra was collected by the on-board  Radial Velocity Spectrometer \citep[RVS,][]{RVS}, the \Gaia\ \gspspec\ module derived the main stellar atmospheric parameters (\T, \g, \meta\ a proxy of [Fe/H]\footnote{The \gspspec\ metallicity parameter \meta\ indeed traces the [Fe/H] abundance ratio \citep[see,][for details]{GSPspecDR3}.}, and the $\alpha$-element abundance enhancement with respect to iron \AF) for up to 5.6 million stars \cite[see][for an extensive description of \gspspec]{GSPspecDR3}.

We here directly estimate supplementary fundamental parameters  from the \gspspec\ parameters and adopt a different method than those described above without relying on any stellar evolution model. The interstellar colour excesses and extinctions towards each \gspspec\ stars are indeed computed from their main atmospheric parameters and predicted colours. Then, after computing the bolometric corrections and adopting complementary \Gaia\ astrometric and photometric data, we estimate the stellar luminosities. Finally, it is straightforward to derive the stellar radii and masses when \T\ and \g are known. All these parameters are thus fully consistent among each other and are free of any assumption of stellar evolution and/or structure. Our work is structured as follows. 
We first describe in Sec.~\ref{Sec:ISM} the computation of the interstellar colour excesses and extinction towards each \gspspec\ star and derive the stellar luminosities, radii, and masses in Sect.~\ref{Sec:LMR}.
Our catalogue content is presented in Sect.~\ref{Sec:Catalog}. 
We explore the effect of possible inaccuracies in the \gspspec\ atmospheric parameters on the interstellar absorption, stellar luminosity, radius, and mass computations in Sect.\ref{Sec:Errors}. The catalogue is then validated by a comparison with literature values in Sect.~\ref{Sec:Valid}, and some application examples are 
discussed in Sect.~\ref{Sec:Discuss}. Our conclusions are finally summarised in Sect.~\ref{Sec:Conclu}.

\section{Interstellar absorption}
\label{Sec:ISM}

To compute stellar luminosities and other related quantities (see Sect.~\ref{Sec:LMR}),
the interstellar extinction in the \Gaia\ band (\AG) towards each \gspspec\ target has to be estimated. For this purpose, we first computed new colour excesses and then extinctions by adopting a method that is fully consistent with the stellar parameters derived by \gspspec, considering only additional information from \Gaia\ two-band photometry. 
Our extinctions can therefore be considered to be purely spectroscopic and have the advantage of not being affected by possible (\T, colour excess) degeneracies, unlike certain photometric methods.

\subsection{Interstellar colour excess}
Since dust reddens stellar colours,
the comparison of intrinsic colours computed from the stellar atmospheric parameters
with observed ones allows us to estimate the colour excess caused by the interstellar medium. This method assumes
that the stars are accurately and precisely parametrised, which is the case for a very large subsample of the \gspspec\ stars. Practically, this exercise was performed for the \Gaia\ $B_P$ and $R_P$ magnitudes, which are available for the whole sample.
The colour excess (or reddening) in the \Gaia\ bands is thus simply expressed by
\begin{equation}
\label{Eq:colour}
E_{(B_P-R_P)} = (B_P -R_P)_{Observed} - (B_P -R_P)_{Intrinsic}.
\end{equation}
The star colours without absorption, $(B_P -R_P)_{Intrinsic}$, were derived with Eq.~1 of \citet[][C21 hereafter, see Eq.~\ref{Eq:Casa}]{Luca21}.
This equation (called $F_{C21}$ hereafter) allowed us to estimate the stellar \T\ based on a polynomial that depends on the observed intrinsic colour, \g\ and \meta.
We recall that a very good agreement in the low-extinction regime between the \gspspec\ parameters and the C21 relation was already confirmed by \citet[][see their Fig.~19]{GSPspecDR3}. This is expected because the two studies relied on similar Solar abundances and stellar atmospheric models, although they were fully independent.

Based on this equation $F_{C21}$ and adopting the \Gaia\ $(B_P -R_P)$ colour, the polynomial coefficients provided in the first line of Tab.~2 of C21, and considering their dwarf and giant separation as a function of the stellar colour, we solved the equation
\begin{equation}
\label{Eq:Casa}
T_{\rm eff} - F_{C21}[\ (B_P -R_P)_{Intrinsic}, {\rm log(g)}, {\rm [M/H]}\ ] = 0 \, , 
\end{equation}
in which we adopted the \gspspec\ \T, \g\ and \meta\ values. 
However, this relation does not take the possible effect on the metallicity of the stellar chemical content into account. We therefore considered in Eq.~\ref{Eq:Casa} a mean metallicity corrected for the stellar \AF\ content, adopting
the relation proposed by \cite{Salaris93}. For the \gspspec\ stars without published \AF\ (11\% of the full catalogue), we assumed that their $\alpha$ content follows the classical Galactic disc relation with metallicity, \AF = 0.0~dex for \meta $\ge$ 0.0~dex,
\AF = +0.4~dex for \meta $\le$ -1.0~dex and
\AF = -0.4 $\times$ \meta\ for -1.0 $\le$ \meta $\le$ 0.0~dex.
These \AF-corrected metallicities were adopted throughout for estimating the other parameters, when necessary.

Furthermore, we recall that the adopted atmospheric parameters were deliberately published uncalibrated by \gspspec, and we favoured the recommended calibrations here as a function of \T , as indicated hereafter. For the surface gravity and metallicity, we adopted the calibrations provided in Appendix~A of \cite{Recio24}, and for \AF\ , we adopted the calibration provided in Table~4 of \cite{GSPspecDR3} as a function of \T. The surface gravity calibration was then slightly improved to further increase the accuracy of \g\ .
For this purpose, a final second-order calibration considering gravity and metallicity effects was applied, based on a comparison with APOGEE/DR17 data \citep{APOGEEDR17}. We refer to \cite{ExoP} for details on this supplementary calibration. We note that the adopted relations hereafter are weakly dependent on the surface gravity (except the stellar mass determination). The values we computed
are thus almost unaffected by possible small calibration inaccuracies.
We finally recall that the recommended calibrations have published validity ranges. For parameters found outside these ranges, we did not extrapolate the 
calibration polynomials and simply adopted the limit correction
values corresponding to the range boundaries.

Eq.~\ref{Eq:Casa} is easily solved numerically by adopting $(B_P -R_P)_{Observed}$ as a first guess,
and then by iterating until a solution is reached at a 10$^{-5}$ level. Typically, the colour excesses are quickly found in a few steps. 
Practically, \EBpRp\ were estimated by solving Eq.~\ref{Eq:Casa} 1000 consecutive times. We
adopted random values for \T, \g, \meta, \AF, and ($B_P - R_P)_{Observed}$ within their error bars and assumed a normal distribution for these uncertainties\footnote{We remind that the \gspspec\ parameter uncertainties mostly reflect the spectra \SNR\ ratios.}. These Monte-Carlo realisations led to rather Gaussian distributions
of \EBpRp. The colour excess values we report (see Sect.~\ref{Sec:Catalog}) are simply the median of the computed distributions.
Their associated uncertainties are then half of the difference between the 84$^{th}$ and 16$^{th}$ quantiles
of the distributions, and hence, they correspond to a 1$\sigma$ uncertainty for a normal distribution.
         
\subsection{Interstellar extinction}
From the previously derived interstellar reddening, we computed the extinction in the \Gaia\ band (\AG). For this purpose, we estimated the coefficients that relate this extinction and the colour excess for every star,
\begin{equation}
k_{TGMA} =  A_G / E(B_P-R_p),
\label{Eq:AG}
\end{equation}
based on the tables provided with the \Gaia\ stellar parameters\footnote{https://www.cosmos.esa.int/web/gaia/dr3-astrophysical-parameter-inference} \citep{Orlagh23}.
These $k_{TGMA}$ coefficients depend on the four atmospheric parameters 
derived by \gspspec: \T, \g, \meta, and \AF. As above, we also considered their associated uncertainties (including those of \EBpRp) to compute 1,000 Monte-Carlo realisations that finally provided the median value of the $A_G$ distributions and their uncertainties.         

\section{Stellar luminosities, radii, and masses}
\label{Sec:LMR}

From the previously estimated interstellar extinction, we computed the stellar luminosity ($L$), radius ($R$), and then, the mass ($M$) for each \gspspec\ star.
It is important to note that the above quantities are independent of any stellar evolutionary models and Galactic priors (except for the adopted distances; see below). 
We first computed the absolute stellar magnitude,
\begin{equation}
M_G = G_{\rm mag} - 5\times {\rm log_{10}}(d) + 5 - A_G \ ,
\end{equation}
where $G_{\rm mag}$ and $d$ (in pc) are the DR3 magnitudes in the \Gaia\ band and the \cite{Coryn21} $geometric$\footnote{These geometric distances were adopted to be consistent with \cite{PVP_Ale}.} stellar distances based on \Gaia/DR3 parallaxes, respectively. The absolute bolometric magnitude is then
\begin{equation}
\label{Eq:Mbol}
M_{\rm bol} = M_G + BC_G \ ,
\end{equation}
where $BC_G$ is the bolometric correction in the $G$ filter, computed according to \cite{Luca18}. This depends on \T, \g, and the $\alpha$-corrected metallicity. The stellar luminosity expressed in Solar units is then easily obtained,
\begin{equation}
{\rm log_{10}(}L / L_\odot) =  -0.4 \times (M_{\rm bol} - M_{\rm bol}^\odot) \ ,
\end{equation}
where $M_{\rm bol}^\odot$=4.74 is the adopted value for the absolute bolometric magnitude of the Sun \cite[IAU 2015 resolution B2,][]{IAU}. The \gspspec\ effective temperature then allowed us to estimate the stellar radius with the Stefan-Boltzmann relation \citep{Stefan1879, Boltzmann1884}, 
\begin{equation}
\label{Eq:Rad}
R / R_\odot = \sqrt{L/ L_\odot} \times ( T_{\rm eff}^\odot / T_{\rm eff})^2  \ ,
\end{equation}
where $R$ is expressed in Solar units, and the effective Solar temperature is fixed at \T$^\odot$=5,777~K. The stellar mass in Solar units was subsequently obtained using the surface gravity,
\begin{equation}
\label{eq:Mstar}
M / M_\odot = (R / R_\odot)^2 \times (g/g_\odot) \,,
\end{equation} where log($g_\odot)$=4.44 with $g$ in cm$^2$/s. 
We point out that precise stellar masses can only be derived from extremely precise $g$ measurements, whereas spectral analysis methods provide \g\ estimates with typical uncertainties of about 0.1-0.2~dex.
Again, we considered the uncertainties associated with all these quantities by performing 1,000 Monte-Carlo realisations, propagating the uncertainties on each atmospheric parameter. As above, we publish the median of the parameter distributions, and the associated uncertainties were estimated based on the 84$^{th}$ and 16$^{th}$ quantiles.

\section{The \gspspec\ catalogue of interstellar absorption and stellar $L$, $R$, and $M$}
\label{Sec:Catalog}

\begin{table}[t]
        \caption{\label{Tab:Cata} Content of the \Gaia\ \gspspec\ catalogue of interstellar reddening, extinctions, stellar luminosities, radii, and masses.}
        \centering
        \begin{tabular}{ll}
        \hline
        Label & Description \\
        \hline
        GDR3Id & $Gaia$ DR3 Identification\\
        \T & Adopted effective temperature (K)\\
        \T\_err & \T\ uncertainty (K)\\
        \g & Adopted surface gravity ($g$ in \gunits)\\
        \g\_err & \g\ uncertainty\\   
        \meta & Adopted metallicity (dex)\\
        \meta\_err & \meta\ uncertainty (dex)\\
        \AF & Adopted enhancement in $\alpha$-elts w.r.t. Fe (dex)\\
        \AF\_err & \AF\ uncertainty (dex)\\        
        E\_BpRp & Colour excess in ($B_P$-$R_P$) (mag.)\\
        E\_BpRp\_err & \EBpRp\ uncertainty (mag.)\\
        A\_G & Extinction in the $G$-band (mag.) \\
        A\_G\_err & A$_G$ uncertainty (mag.) \\
        L & Stellar luminosity (in $L_\odot$)\\
        L\_err &  $L$ uncertainty (in $L_\odot$) \\
        R &  Stellar radius (in $R_\odot$) \\
        R\_err &  $R$  uncertainty (in $R_\odot$)  \\
        M &  Stellar mass (in $M_\odot$) \\
        M\_err &   $M$ uncertainty (in $M_\odot$) \\
        Flag\_Param &  QF associated to the \gspspec\ parameters \\
                    &  (0/1/2) = (High/Good/Moderate)-Quality\\
        Flag\_Ext &  QF associated to the extinction calculation\\
                  &  (0/1/2) = (High/Good/Moderate)-Quality\\
        Flag\_LRM &  QF associated to the $L$, $R$, and $M$ derivation\\                  &  (0/1/2) = (Very high/High/Medium)-Quality\\
                  &  (9) = Low-Quality astrometric parameters \\
                  &  \hspace{0.7cm}      and/or $LRM$ filtered out (see text)\\ 
         \hline
        \end{tabular}
        \tablefoot{The full version of this table is available at CDS.}
\end{table}

The computed colour excesses, extinctions, luminosities, radii, and masses together with their associated uncertainties are reported in an electronic table (see Table~\ref{Tab:Cata} for its content) for each \gspspec\ star, when available.
We also include in the table the calibrated atmospheric parameter values that were adopted 
in this paper.

\subsection{Parameter accuracy and quality flags}
Together with these parameters, we also publish three quality flags ($QF$, best value=0) that
allow one to select the best derived quantities. Except for the first flag, the two additional quality flags are specific to the derivation of the new parameters derived here. These new $QF$s are defined as described below:

\begin{itemize}
\item The goal of $Flag_{\rm Param}$ is to summarise the flag chain of \gspspec\ into a single value for the use of the new data presented here. We recall that the 13 first \gspspec\ $QF$s refer to the quality of the \T, \g, and \meta\ estimates \citep[see for more details Sect.8 of][]{GSPspecDR3},
mostly depending on the quality of the input RVS spectra. The best parametrised $(\sim$1.9~million stars, i.e. $\sim$40\% of the $\sim$4.6 million included in this catalogue) for which all their first 13 \gspspec\ flags are equal to 0 therefore have $Flag_{\rm Param}$=0. Moreover, $Flag_{\rm Param}$=1 (2) refers to stars for which at least one of these 13 first QFs is equal to 1 (2), and this affects $\sim$35\% stars of the whole sample (2; $25$\% stars). We note that about $20$\% of all the \gspspec\ stars were rejected from our catalogue because the quality of their parameters was too low, as 
revealed by one of their \gspspec\ QFs being equal to 9 \cite[some of them being also
KM-giant flagged stars, see][]{GSPspecDR3}. 

\item $Flag_{\rm Ext}$ allows one to select targets with the best interstellar absorption estimates. It should therefore be considered when dealing with \EBpRp\ and/or \AG, but also with $L$, $R$, or $M$ because these quantities rely directly 
on the interstellar extinction estimate. 
It can take values of 0, 1, or 2 (from high quality to moderate quality, respectively) and was determined 
by considering the accuracy reached when numerically solving Eq.~\ref{Eq:Casa}.
Stars whose $(B_P -R_P)_{Observed}$ colour was found to be outside the validity range of the $F_{C21}$ function in Eq.~\ref{Eq:Casa} were filtered out.
Typically, this occurred for stars that are too cool.
The very few stars for which Eq.~\ref{Eq:Casa} cannot be numerically solved were also filtered out. We proceeded similarly for stars whose effective temperature was outside the recommended range by C21 (i.e. cooler than 3600~K or hotter than 8500~K) and for those whose \g\ and/or \meta\ was extrapolated with respect to the ranges of the synthetic spectrum reference grid adopted by \gspspec. In summary, as a consequence of these rejection criteria (including those related to $Flag_{\rm Param}$), the total number of stars in the catalogue is $\sim$4.6 million. 

\item $Flag_{\rm LRM}$ assesses the quality of the $L$, $R$, and  $M$ computation. Basically, this flag was estimated by propagating 
the $Flag_{\rm Ext}$ value and considering whether the \g\ and \meta\ values adopted for the computation  of the bolometric correction are well found within the validity ranges recommended by \cite{Luca18}.
Four different values are provided: 0, 1, and 2
for very high, high, and medium-quality determinations; and determination with a quality that is too low
are identified by $Flag_{\rm LRM}$=9. 
Moreover, no $LRM$ values are published for the $\sim$1.4\% stars with $Flag_{\rm LRM}$=9, and when  Eq.~\ref{Eq:Casa} cannot be solved and/or when no $BC_G$ were estimated because their stellar atmospheric parameters
are found to lie outside the validity ranges provided by \cite{Luca18}.
In contrast, some stars with $Flag_{\rm LRM}$=9 have published $LRM$s. They have possible \Gaia\ astrometric issues that might affect their distance (and thus, $L$) estimates. These stars either have a \Gaia\ astrometric $ruwe$ parameter larger than 1.4 or a $fidelity$ factor lower than 0.5, following \cite{Rybizki22}; both of these indicate a poor astrometric solution. Possible non-single stars are also found in this category and were identified by considering the \Gaia\ keywords $duplicated\_source$ and $non\_single\_star$. In total, all these astrometric criteria affect about 20\% of all the \gspspec\ stars. We finally note that some variable stars might be included in our catalogue. Future users should check their parameters since the \gspspec\ analysis might have been affected by the stacked-epoch RVS spectra.
\end{itemize}

\begin{figure}[t]
    \centering
    \includegraphics[width=0.45\textwidth]{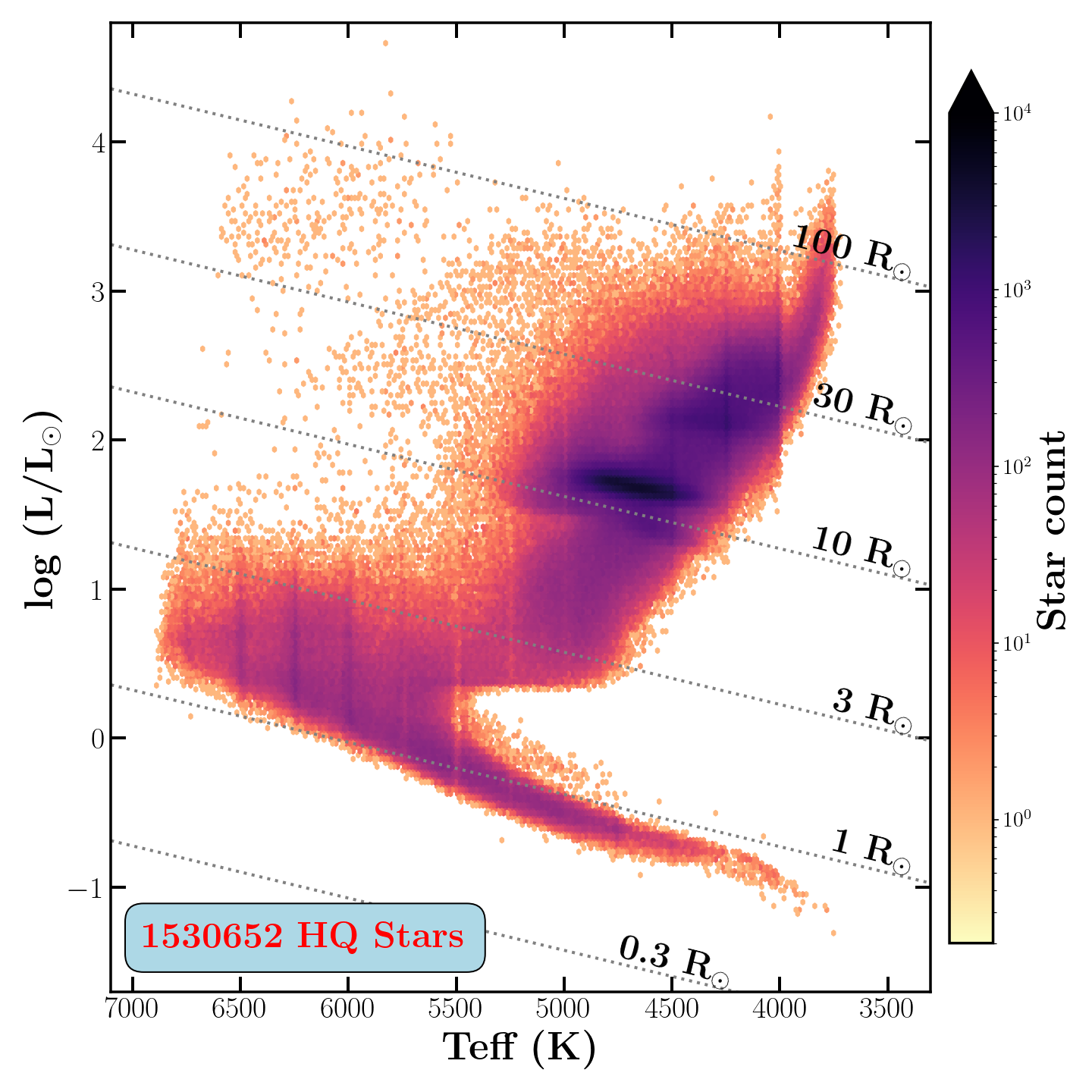} 
    \caption{Stellar luminosity vs. effective temperature diagram for the high-quality subsample (see text for its definition), colour-coded by the stellar count. The dotted lines represent the iso-radius relations.} 
    \label{Fig:L-Teff}
\end{figure}
\begin{figure*}[t]
    \centering
    \includegraphics[width=0.45\textwidth]{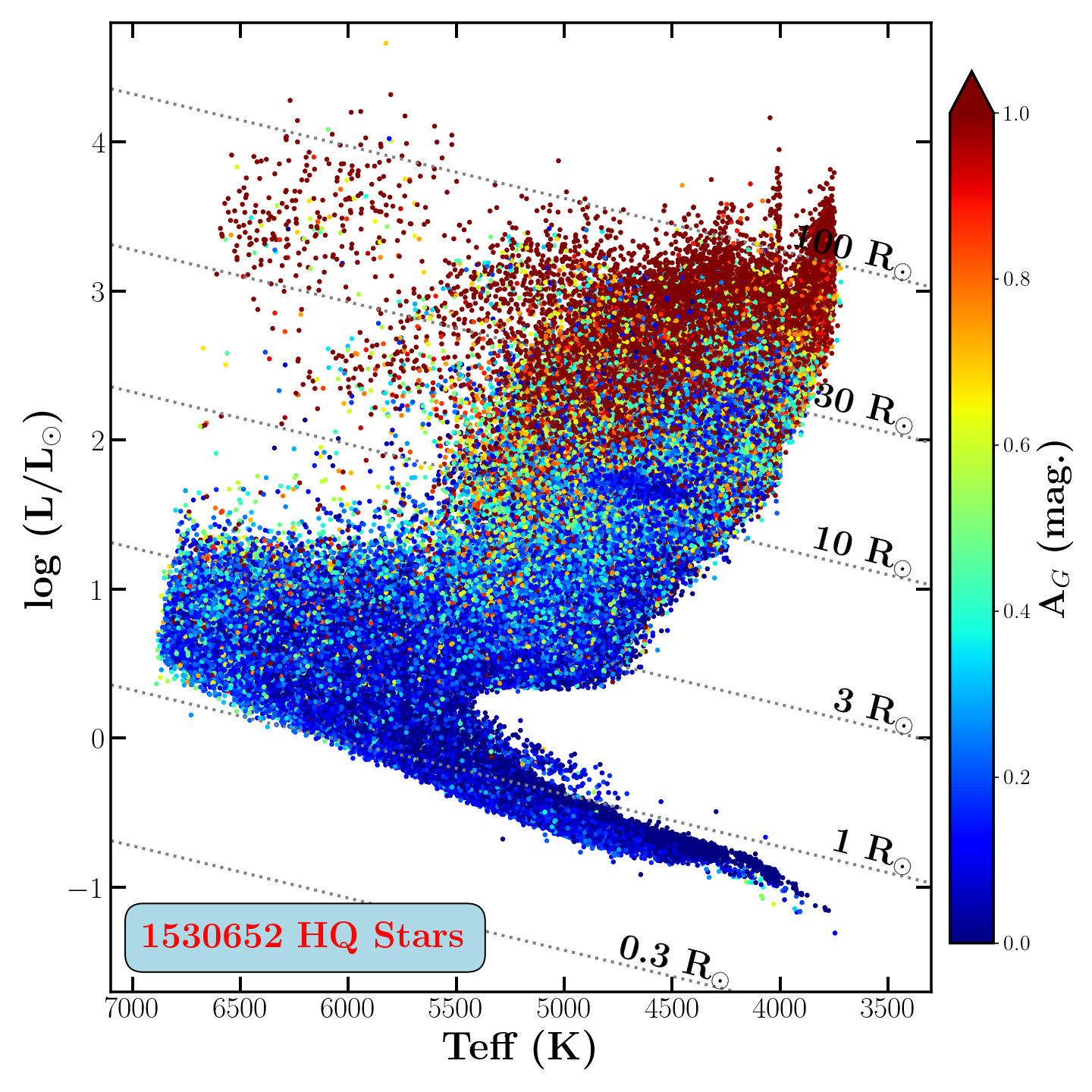} 
    \includegraphics[width=0.45\textwidth]{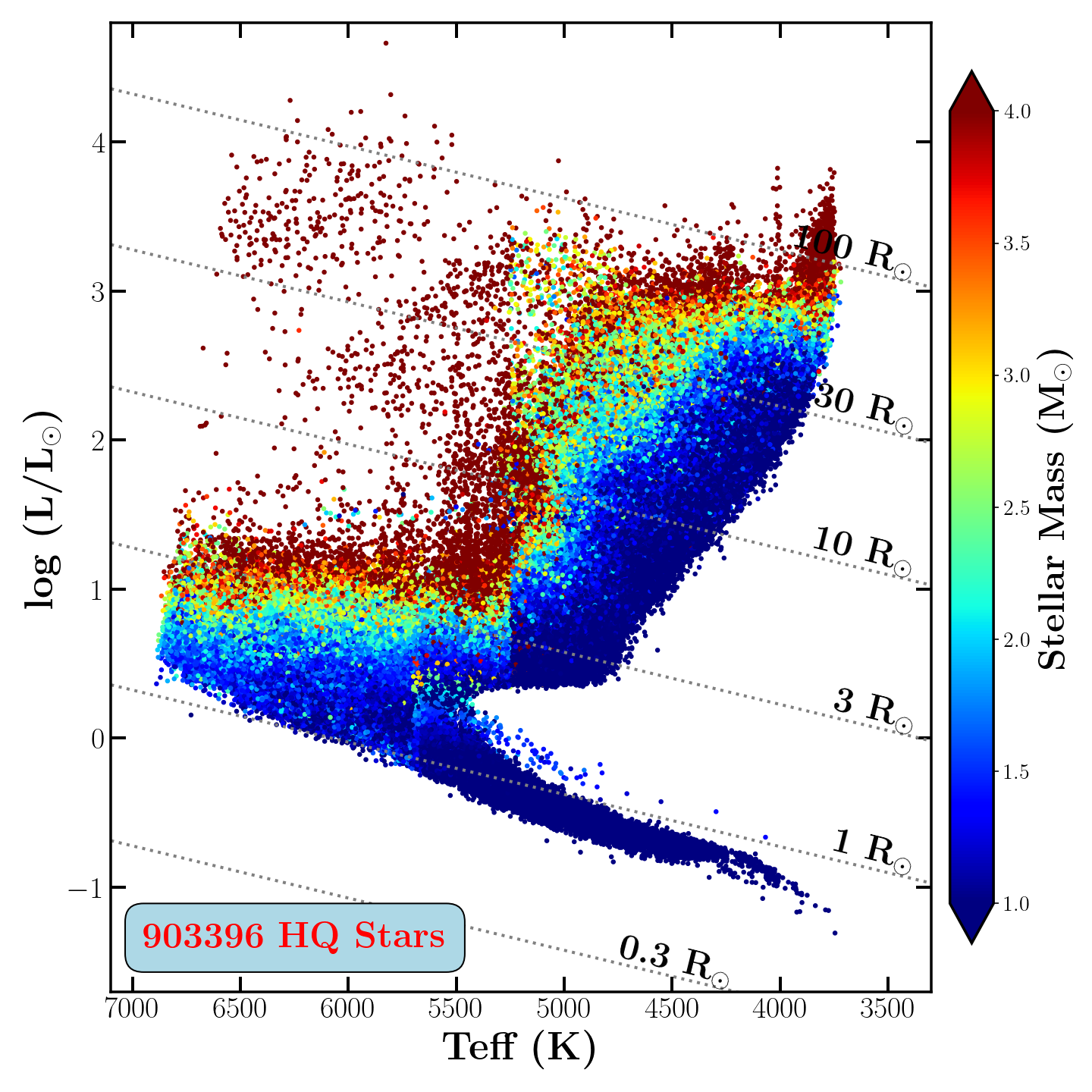} 
    \caption{Same as Fig.\ref{Fig:L-Teff}, but colour-coded by interstellar absorption and stellar mass (from left to right).} 
    \label{Fig:L-Teff2}
\end{figure*}

Moreover, a final check was specifically performed for the derived masses since it is well known that for some stars, the surface gravity might be difficult to derive accurately from their spectra. Because $M$ varies linearly with $g$ (and not with \g, as indicated in Eq.~\ref{eq:Mstar}), rather large uncertainties in \g\ lead to huge $M$ uncertainties. For instance, a 0.3~dex in \g\ uncertainty corresponds to a relative mass error larger than a factor of 2 (i.e. $\sim$100\% for $M$).
This should be considered when building working samples for specific scientific applications (see below). Furthermore, we also searched for possible large inconsistencies between the \gspspec\ \g\ and a luminosity-based surface gravity estimate,
as was done in \cite{GamDor}. These were simply estimated by computing \g$_{\rm Lum}$ = log($g_\odot)$ + log($M$) - log($L$) + 4$\times$log(\T$^\odot$), adopting a very wide range of possible masses. About $10^4$ stars with likely poor \gspspec\ gravities were then identified when the difference between the spectroscopic and luminosity-based surface gravities was larger than 2~dex. Their masses were rejected from the catalogue, but the other derived parameters remained included because they depend very weakly on \g.

Finally, in addition to these quality flags, we also recommend to consider the reported uncertainties associated with the different parameters in order to define high-quality subsamples that are especially adapted for specific applications. For instance, stellar radii and masses for stars with a large luminosity uncertainty (e.g. $\ga$30\%) should be considered with caution. Moreover, the mass of stars with a surface gravity uncertainty larger than $\sim$0.3~dex could be filtered out.
We also encourage future users to also adopt the quality flag system of \cite{GSPspecDR3} to more precisely select stars with optimal \T, \g\, and \meta\ derivations instead of considering $Flag_{\rm Param}$. The spectra \SNR\ and the \gspspec\ $\chi^2$ quality-fit parameter could also be considered. \\

\subsection{Catalogue content and typical uncertainties}
In the catalogue presented in Table~\ref{Tab:Cata}, we finally publish the colour excess and extinction for 4.6 million stars, that is, 82\% of the whole \gspspec\ catalogue. Luminosity 
and radius are provided for 99.7\% of them, and the mass is provided for 99.5\%.
In this catalogue, 62\% of the stars have their three $QF \le 1$, meaning a good-quality derivation, and 1.52 million stars (33\%) have very high quality parameters with $Flag_{\rm Param}$=$Flag_{\rm Redd}$=$Flag_{\rm LRM}$=0. 
As an illustration of the catalogue content, we show in Fig.~\ref{Fig:L-Teff}\footnote{In this figure, the overdensity features seen at some \T\ reference grid points are caused by overfitting patterns produced by the \gspspec\ algorithm in the high-S/N regime \citep[see][for more details]{GSPspecDR3}.} and \ref{Fig:L-Teff2} diagrams of the luminosity versus effective temperature for a high-quality ($HQ$) subsample, colour-coded by stellar count, the extinction, and the stellar mass. 
These 1.53 million $HQ$ stars (about one-third of the total sample) were selected by applying the following criteria: ($Flag_{\rm Param}$=0) \& ($Flag_{\rm Redd} \le 1$) \& ($Flag_{\rm LRM} \le 1$), plus a relative error on the luminosity lower than 30\%. Only for exploring the stellar masses, we then restricted this $HQ$ sample to stars with \g$_{\rm err}<0.1$ (i.e. equivalent to an error on $g$ lower than 25\%), leading to about 900,000 stars with a rather well-defined mass (20\% of the total sample). The mean uncertainties associated with this $HQ$ sample are equal to 0.02 and 0.03~mag for \EBpRp\ and \AG, and 6\%, 3\%, and 22\% for the mean relative uncertainties on $L$, $M$, and $R$, respectively. The largest relative errors on $R$ and $M$ are equal to 16 and 40\%.
This $HQ$ sample is also characterised by (i) $\sim$24,000 stars with a luminosity lower than 0.5~$L_\odot$ and $\sim$22,000 high-luminosity stars ($L>500~L_\odot$), 
(ii)  $\sim$20,000 dwarfs with a radius smaller than 0.8~$R_\odot$ and $\sim$500 giants defined by $R>100~R_\odot$ (only two have $R>200~R_\odot$), and (iii) $\sim$9,000 stars are less massive than 0.6~$M_\odot$ and $\sim$2,300 are more massive than 10~$M_\odot$ (168 with $M>50~M_\odot$). 
More exotic stars should be explored outside the $HQ$ sample by relaxing the quality flags.
This Fig.~\ref{Fig:L-Teff} shows that the different evolutionary stages of stars of any masses are well represented. The majority of the most reddened stars are the most luminous, massive, and hence, distant stars belonging to the Galactic halo, whereas a few other massive stars are much closer but deeply embedded in the thin disc. In contrast, the main sequence is highly dominated by low-mass dwarfs with low extinction since they are not very distant. Finally, at any stage, a correlation between
$L$ and $M$ is clearly seen.

\section{Effect of the atmospheric parameter accuracy on the derived \AG, $L$, $R$, and $M$}
\label{Sec:Errors}
\begin{table*}[h]
    \centering
        \caption{Effect of possible atmospheric parameter inaccuracies on the quantities derived here.}
    \begin{tabular}{llccccc}
        \hline
         & & Main Sequence &  Turn-Off & Lower RGB &  Clump &  AGB\\  
         & & (5000~K, 4.5) &  (6500~K, 3.5) & (4800~K, 3.0) &  (4500~K, 2.5) &  (4000~K, 1.5)\\    
        \hline  
        & & & & & & \\ 
       $\Delta$\T=100~K &  $\Delta$ (\AG)& 0.07~mag & 0.05~mag & 0.09~mag  & 0.11~mag  & 0.18~mag  \\
                       &  $\Delta (L, R,  M)$ & (5.1, 1.5, 2.9)\% & (4.8, 0.7, 1.6)\% & (5.9, 1.1, 2.3)\% &(6.8, 1.1, 2.3)\% &(11., 0.2, 0.3)\% \\
        & & & & & & \\    
       $\Delta$\g=0.25~dex &  $\Delta$ (\AG)& 0.01~mag & 0.02~mag & 0.02~mag  & 0.01~mag  & 0.01~mag  \\
                       &  $\Delta (L, R,  M)$ & (0.8, 0.4, 77.)\% & (1.1, 0.6, 75)\% & (1.5, 0.7, 75.)\% &(0.8, 0.4, 77.)\% &(1.1, 0.5, 76)\% \\
        & & & & & & \\   
        $\Delta$\meta=0.1~dex &  $\Delta$ (\AG)& 0.01~mag & 0.01~mag & 0.01~mag  & 0.01~mag  & 0.01~mag  \\
                       &  $\Delta (L, R,  M)$ & (1.1, 0.6, 1.2)\% & (0.7, 0.3, 0.9)\% & (0.7, 0.3, 0.7)\% &(0.6, 0.3, 0.6)\% &(0.8, 0.4, 0.8)\% \\
        & & & & & & \\   
       $\Delta$\AF=0.1~dex &  $\Delta$ (\AG)& 0.01~mag & 0.01~mag & 0.01~mag  & 0.01~mag  & 0.01~mag  \\
                       &  $\Delta (L, R,  M)$ & (0.7, 0.4, 0.6)\% & (0.4, 0.2, 0.6)\% & (0.4, 0.2, 0.5)\% &(0.5, 0.3, 0.3)\% &(0.6, 0.3, 0.9)\% \\
        & & & & & & \\           
        \hline
    \end{tabular}
    \tablefoot{These test stars are typical of different evolutionary stages. Their adopted effective temperatures and surface gravities are provided in the second line. Solar \meta\ and \AF\ are adopted for all of them.}
    \label{Tab:inaccurate}
\end{table*}

The uncertainties on \AG, $L$, $R$, and $M$ provided in Table~\ref{Tab:Cata}
were directly estimated from the uncertainties of the atmospheric parameters (\T, \g, \meta, and \AF) reported in the \gspspec\ analysis. We recall that these uncertainties result from the RVS spectrum noise and are thus only related to the precision of the parameter derivation. We next explore the effect of the accuracy of the atmospheric parameter derivation on the new quantities we derived. 

For this purpose, we simply (i) considered some test stars  typical of the lower main sequence, the turn-off, the lower Red Giant Branch (RGB), the clump of the Horizontal Branch, and the Asymptotic Giant Branch (AGB; see Table~\ref{Tab:inaccurate} for their parameter values), (ii) varied their atmospheric parameter as if they were inaccurate, (iii) rederived their \AG\footnote{In Table~\ref{Tab:inaccurate}, we provide only the effect on \AG\ since it is identical as the one on \EBpRp.}, $L$, $R$, and $M$, and finally quantified the effect of these inaccuracies. For each test star, we adopted Solar metallicity and \AF, and we considered no uncertainties on the distance and magnitudes. The inaccuracies we explored for \T, \g, \meta, and \AF\ were 100~K, 0.25~dex, 0.1~dex, and 0.1~dex, respectively. These values are typical of those associated with spectroscopic surveys.

First, as shown in Table~\ref{Tab:inaccurate}, possible inaccuracies in \meta\ and \AF\  do not affect any of our parameters at all, regardless of the stellar type, because these two atmospheric parameters only act in Eq.~\ref{Eq:Casa}  and \ \ref{Eq:AG} and on the bolometric correction, and the dependence of these relations on these parameters is rather weak for an inaccuracy of 0.1~dex. Similarly, and for the same reasons, the inaccuracy in \g\ does not affect the derivation of \AG, and $L$ and $R$ are safely estimated (with an uncertainty of $\sim$1\%) regardless of the stellar type. 
In contrast, an inaccurate \T\ by 100~K can lead to rather strong variation in the absorption of 0.05 up to 0.2~mag, and thus, to luminosity variations of typically $\sim$5-10\%. The stellar mass and radius are less affected by $\Delta$\T, however, with a possible variation of up to $\sim$3\%. This effect is stronger for the giant stars.
It is simply caused by a significant change in the theoretical stellar colour when \T\ is varied (derived from Eq.~\ref{Eq:Casa}), which then affects Eq.~\ref{Eq:colour}, and, as shown in Eq.\ref{Eq:Rad} and \ref{eq:Mstar}, $\Delta R$ and $\Delta M$ vary mostly as $\Delta$\T/\T$^{-3}$. 
Finally, and as expected when examining Eq.~\ref{eq:Mstar}, the mass of all star types is strongly modified by about $\sim$75\% for an inaccuracy in \g\ of only 0.25~dex. 

In summary, interstellar extinction and stellar luminosities are accurately derived for stars with quite accurate effective temperatures. Similarly,
we also recommend 
adoption of the masses published here for stars with very accurate stellar gravities. We excluded masses from our $HQ$ sample derived from \g\ with uncertainties larger than 0.1~dex for this reason (among others; see the previous section). We also refer to Sect.~\ref{Sect:summary} and the validations with asteroseismic masses, which confirm
this behaviour.

\section{Validation with the literature}
\label{Sec:Valid}

In this section, the \Gaia\ \gspspec\ catalogue of interstellar reddening and stellar radii and masses is compared to literature values
for validation purposes. We first summarise some of our previous comparisons of \EBpRp, $R$, and $M$ with published values, and we then focus on more global validations of our stellar radii through interferometric angular diameters and of 
our $R$ and $M$ exploiting asteroseismic data. We note that we did not consider stellar radii derived from surface brightness-colour relations for these validations because they require specific calibrations
that might be affected by distance and reddening uncertainties, in contrast to asteroseismic determinations \cite[see, for instance, the series of articles from][]{Valle24}.

\subsection{Previously published comparisons}
\subsubsection{Interstellar reddening}
The computed colour excesses towards each \gspspec\ star have been extensively discussed and compared to literature values by \cite{Marie26}. We refer to this paper for a detailed validation of the derived 2D and 3D \gspspec\ extinction maps. In brief, a very good agreement with the literature was found, and an excellent spatial correlation with already known interstellar medium structures such as molecular clouds and spiral arms was retrieved. This study therefore confirms the high quality of the interstellar reddening values in the present catalogue.

\subsubsection{Radii and masses of specific stars}
To validate the stellar radii and masses we estimated, we first referred to \citet{Recio24, ExoP, Camila} who validated these quantities for specific
stellar types, but for a rather small number of stars. More global comparisons performed for larger numbers of stars are presented in Sect.~\ref{Sect:GlobalValid}.
First, \cite{Recio24} considered a sample of \gspspec\ stars with extremely precise atmospheric parameters. They found that our spectroscopic masses and those derived from combined Kepler and APOGEE data agree very well \citep[APOKASC2,][]{APOKASC2} for stars with similar \g\ values in the two catalogues. These authors found
a small bias of about 0.03~$M_\odot$, associated with a dispersion of 0.22~$M_\odot$ (see their Fig.~4). Then, \citet[][see their Appendix~A]{ExoP} explored the stellar radii and masses for the exoplanet host stars in our catalogue and compared them with values derived from asteroseismic and spectroscopic data. Again, the agreement with the literature was found to be very good, with no bias and a dispersion of 4\% for $R$ and a bias of 4\% associated with a dispersion of 17\% for $M$. Finally, we also refer to \cite{Camila}, who validated our masses for extreme cases (red giant binaries) and also found a satisfactory agreement with published values.\\

\subsection{Global validation}
\label{Sect:GlobalValid}
In the following, we proceed with a more global validation of our whole catalogue, for which we not only consider
some specific applications as described above. In order to facilitate the comparison, we only considered homogeneous reference catalogues of stellar radius and mass determinations, based on direct technics (i.e. the least model dependent), such as interferometry and asteroseismology. \\

\subsubsection{Interferometric angular diameters}
\begin{figure}[]
    \centering
   \includegraphics[width=0.45\textwidth]{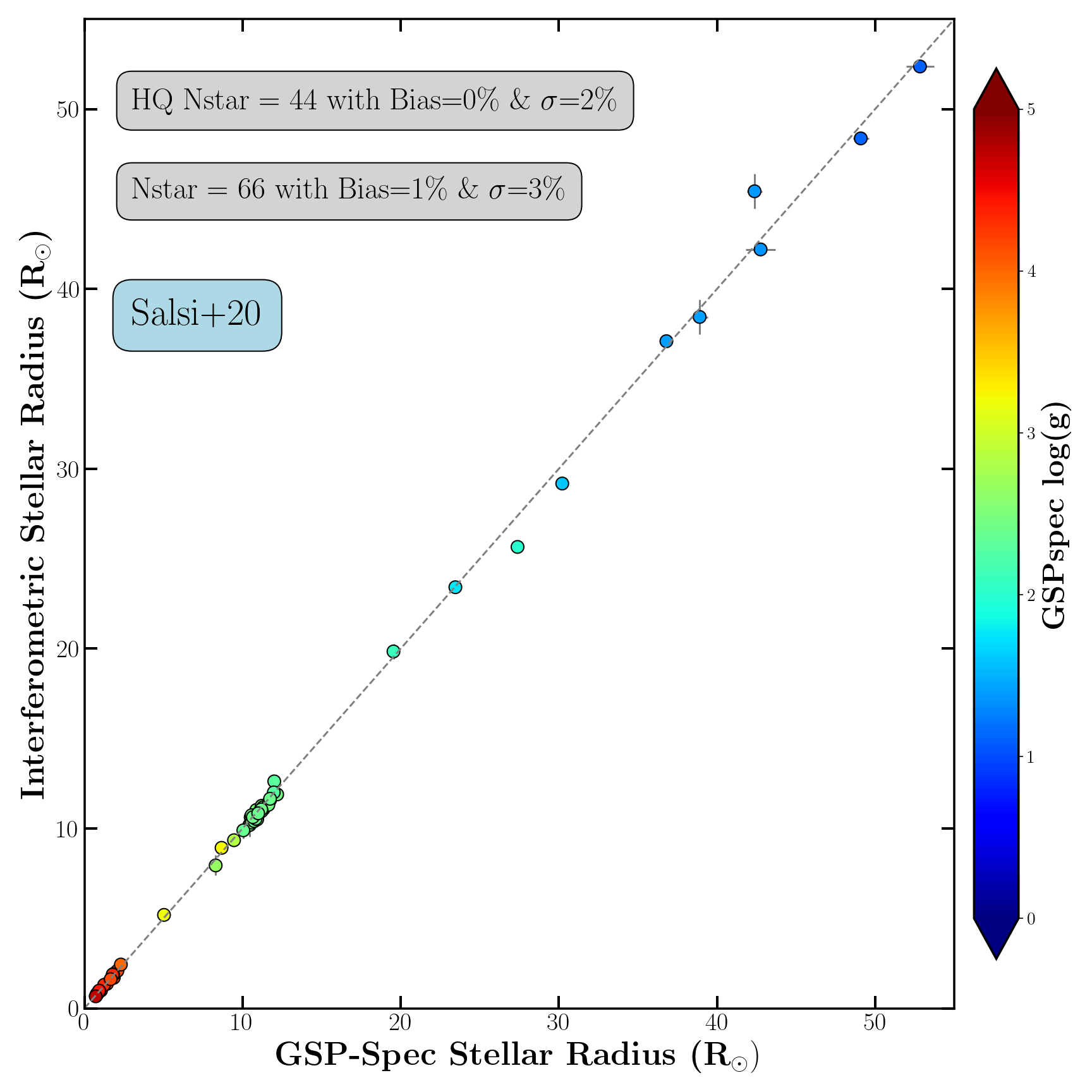} 
    \caption{Comparison between the measured interferometric stellar radii from \cite{Salsi20} and our values.  
    The 1:1 relation is shown as a dashed line.  Most error bars are too small to be visible. The colour-coding represents the \gspspec\ surface gravity, which confirms the dwarf or giant nature of the S20 stars. When our 44 $HQ$ stars are considered alone, the bias becomes null and the dispersion improves slightly.}
    \label{Fig:Salsi}
\end{figure}

Our stellar radii can be directly compared to interferometric observations that provide stellar angular diameters, which, combined with the \cite{Coryn21} distances we adopted, allow us to obtain a completely independent measurement of the stellar radius. For this comparison, we considered angular diameters of dwarfs and giants from the catalogue of \citet[][S20 hereafter]{Salsi20}, 
who implemented a selection of reliable 
interferometric measurements from the
JMMC Measured Stellar Diameters Catalog
\citep{DuvertJMMC}.  This catalogue contains 127 stars with a high-quality diameter measurement (adopting their Rej\_Keep flag), 66 of which are also found in our catalogue. Fig.~\ref{Fig:Salsi} shows a comparison between these interferometric stellar radii and our spectroscopic ones. An impressive agreement is found: the mean relative difference (computed as 
($R_{\rm \gspspec}$ - $R_{\rm S20}$)/$R_{\rm S20}$) is equal to 1\% with a dispersion of only 3\%. Moreover, this bias becomes null when we only consider the 44 stars belonging to our $HQ$ sample, and the dispersion is reduced to 2\%s.
The agreement between these two completely independent radius determinations is therefore excellent and validates the method we adopted, that is, not only the radius determination, but also the different previous steps described in Sect.~\ref{Sec:ISM} \& \ref{Sec:LMR}: the derivation of the interstellar extinction, the bolometric correction, and the luminosity from the \gspspec\ parameters.\\

\subsubsection{Asteroseismic radii and masses}
\begin{figure*}[]
    \centering
    \includegraphics[width=0.42\textwidth]{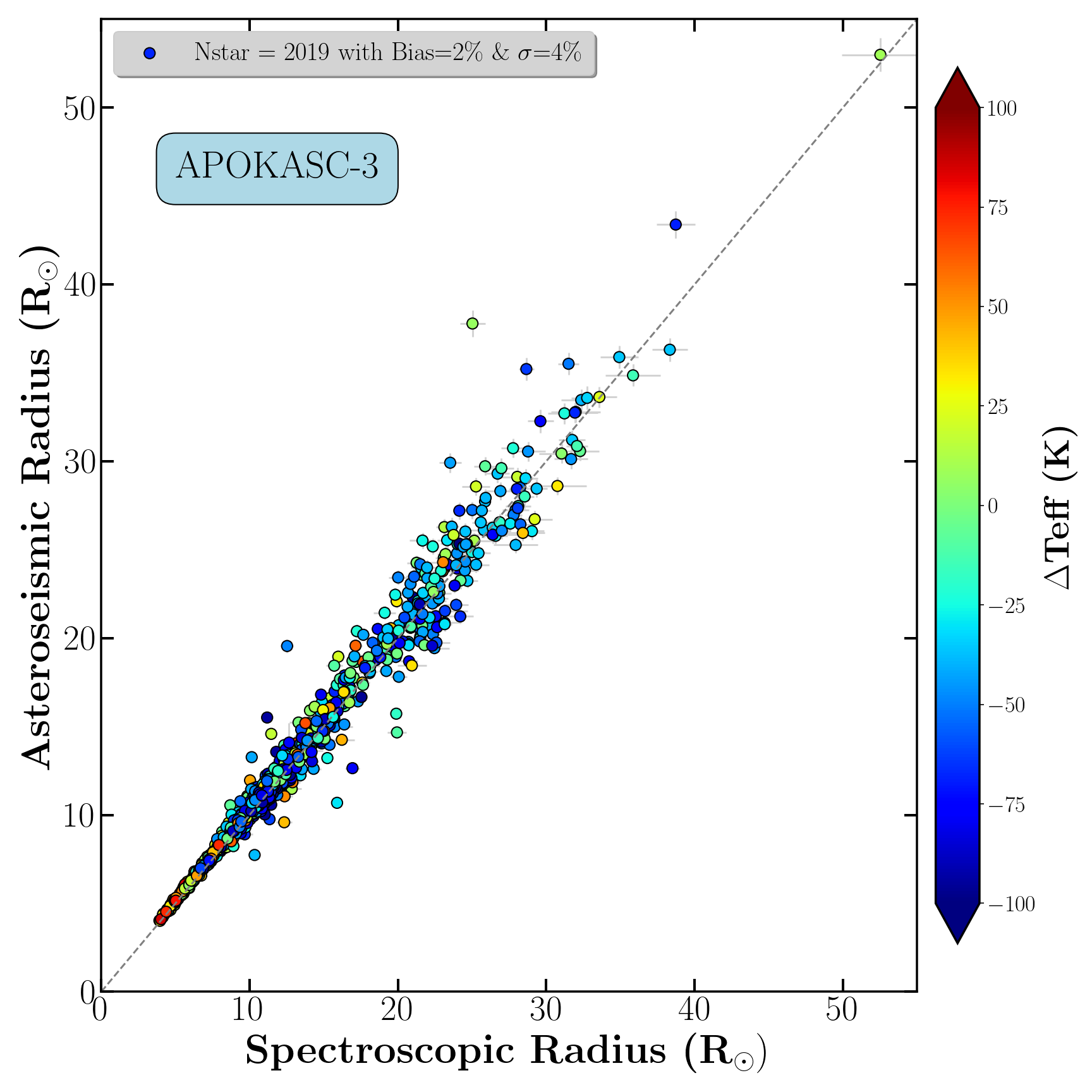} 
    \includegraphics[width=0.42\textwidth]{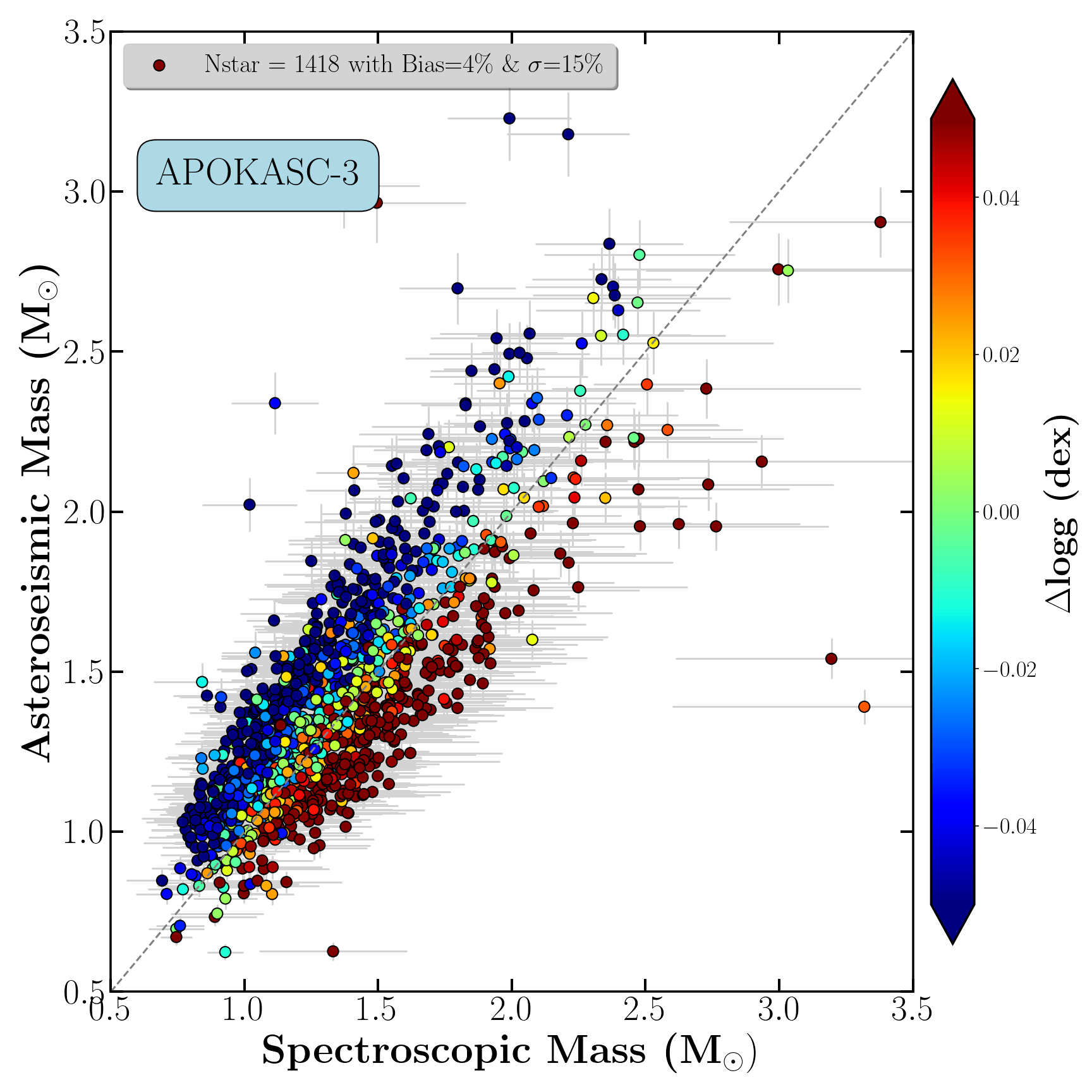}
    \includegraphics[width=0.42\textwidth]{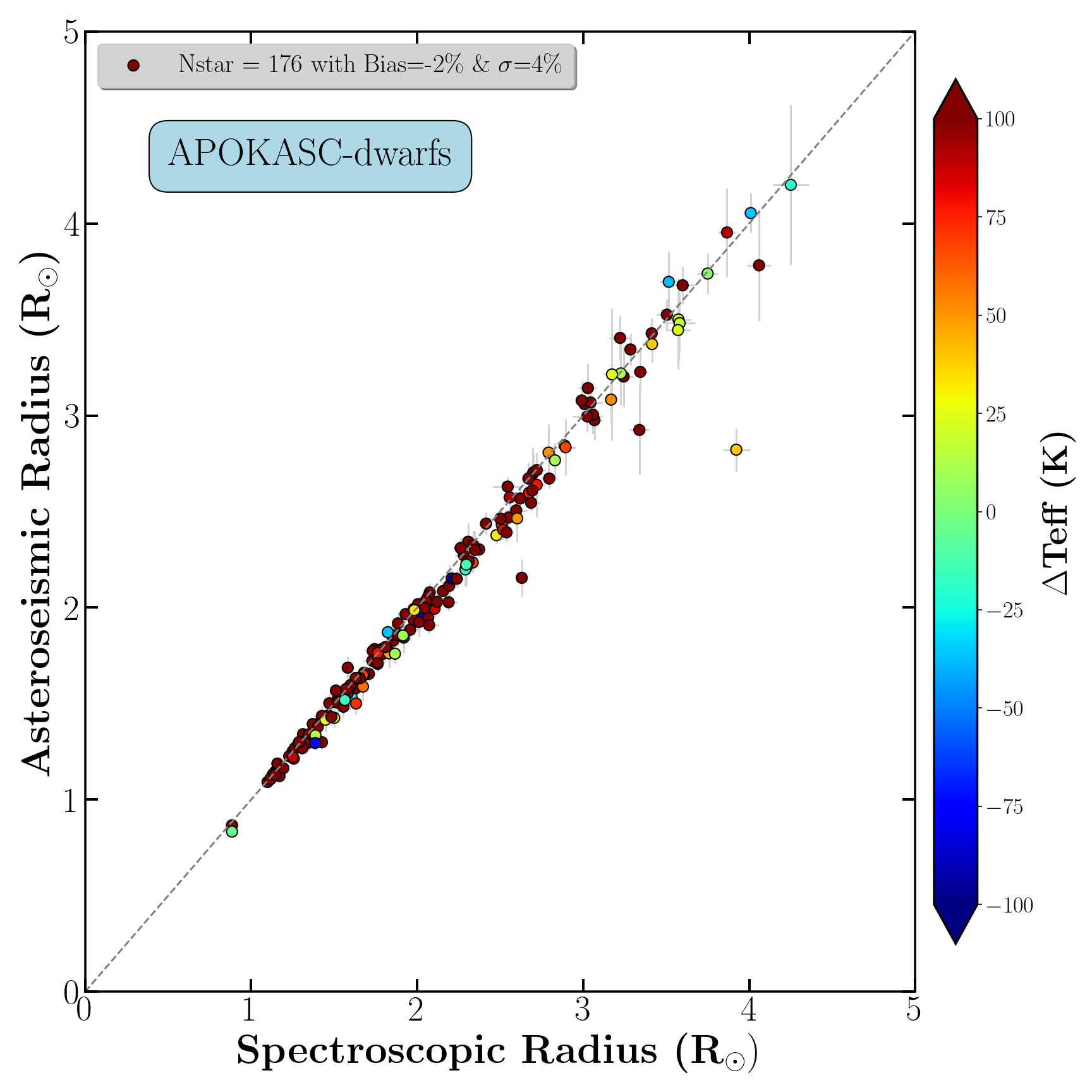} 
    \includegraphics[width=0.42\textwidth]{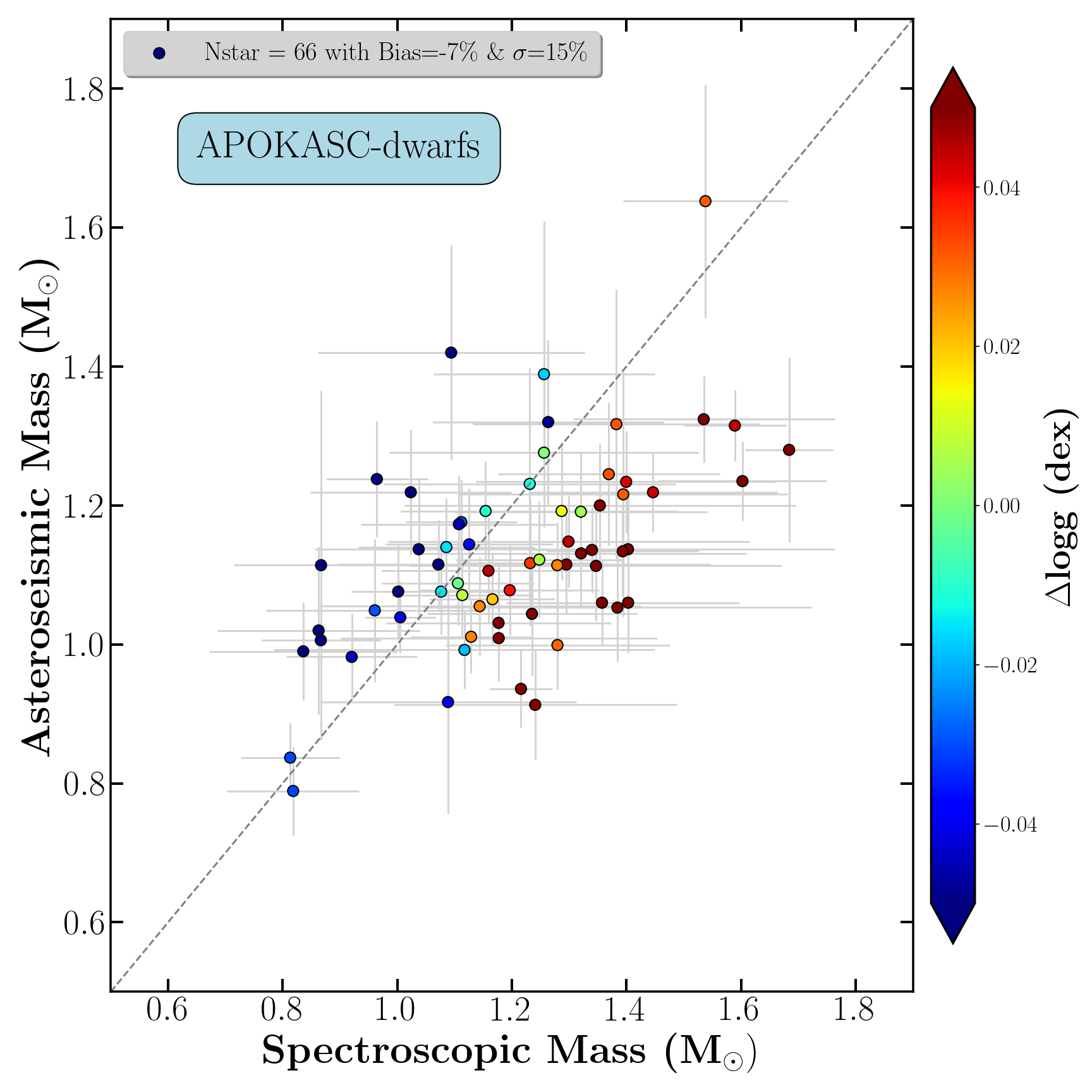}     
    \includegraphics[width=0.42\textwidth]{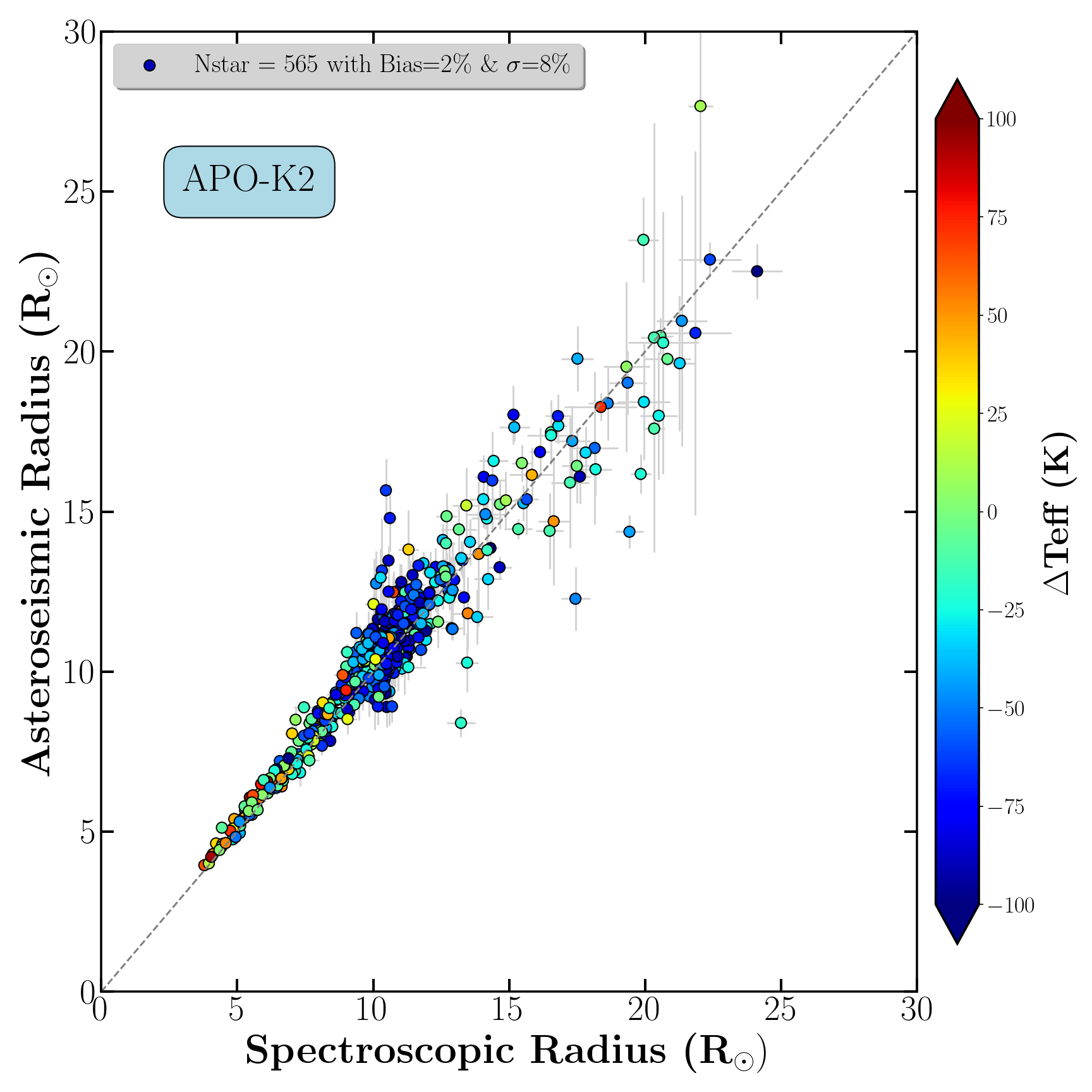} 
    \includegraphics[width=0.42\textwidth]{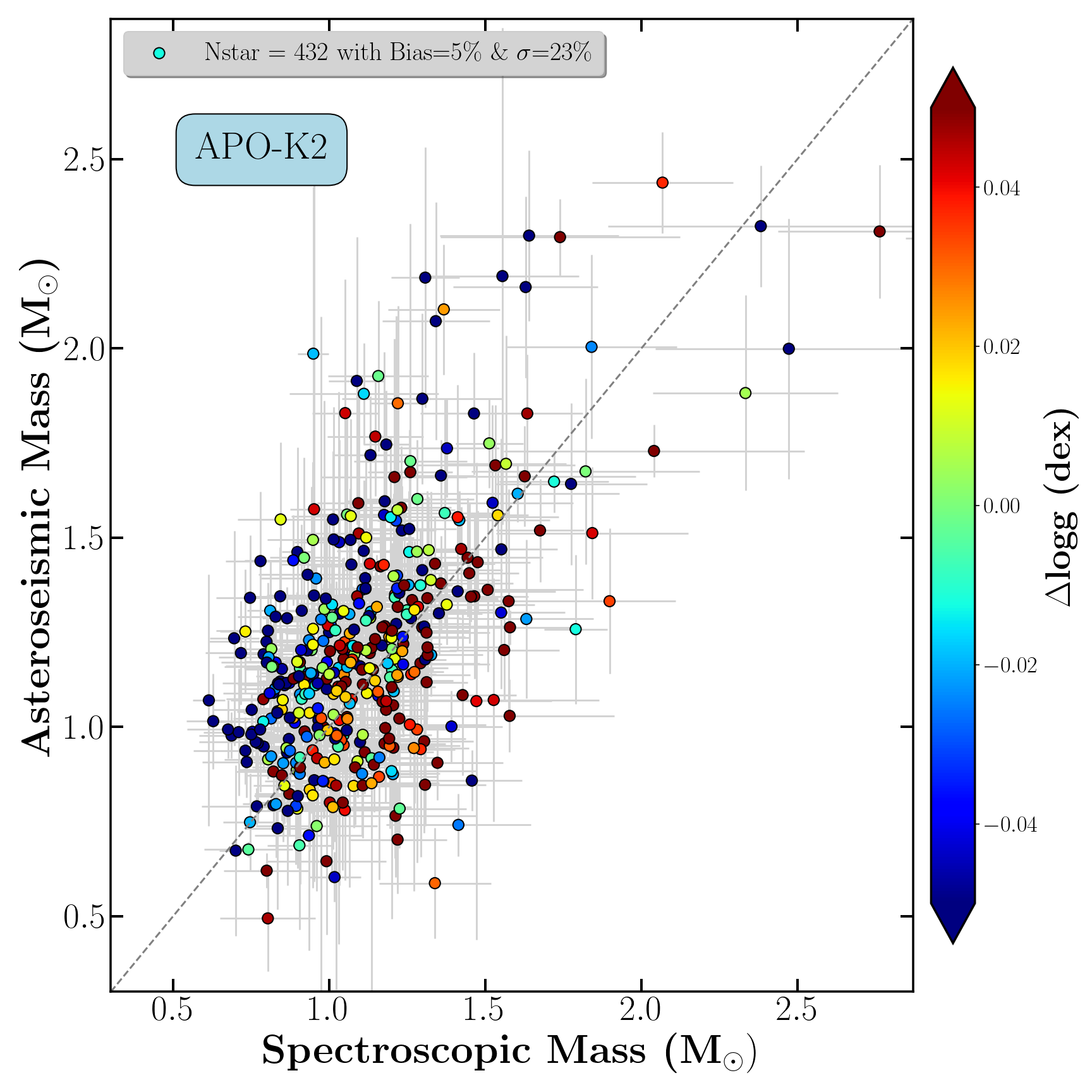}     
    \caption{Comparison between the \gspspec\ spectroscopic stellar radii (left panels) and masses (right panels) and those derived based on asteroseismic data by \citet{APOKASC3,APOKASCnaines,Zinn22}, shown in the top, middle, and bottom panels, respectively. The 1:1 relation is shown as a dashed grey line. The colour-codes represent the difference in \T\ and \g\ between the different samples in the sense (\gspspec\ - literature), left and right panels, respectively. The mean relative differences and associated standard deviations indicated in each panel are in the sense of (\gspspec\ - literature)/literature (see the associated text for more details).} 
    \label{Fig:ValidMR}
\end{figure*}

We compared our best derived stellar radii and masses ($HQ$ subsample)
to completely independent estimates based on asteroseismic observations. For this purpose, we considered the radii and masses of giant stars derived by the APOKASC-3 and APO-K2 studies \cite[respectively]{APOKASC3, APOK2}
and the APOKASC values of \cite{APOKASCnaines} for dwarfs and subgiants.
The APOKASC and APO-K2 studies relied on Kepler  \citep{Kepler} and K2 datasets, respectively, which differ
in several aspects: asteroseismic data analysis, target selection, shorter K2 observations, and so on. So, differences in radius and mass are expected between these studies (see an illustration below).
Moreover, we recall that the asteroseismic radius depends on \T$^{1/2}$ and that our spectroscopic radius is estimated from the square root of the luminosity and \T$^2$ (see Eq.\ref{Eq:Rad}). Similarly, the asteroseismic mass is related to \T$^{3/2}$ and the spectroscopic mass to \g\ (Eq.~\ref{eq:Mstar}). In the following validations, we therefore restricted the comparison samples to stars with a difference in \T\ and \g\ smaller than 100~K and 0.1~dex, respectively, between the \gspspec\ and asteroseismic\footnote{When available, we adopted their asteroseismic \g\ and not their spectroscopic values} catalogues. \\

\noindent APOKASC:
The third version of the APOKASC catalogue (APOKASC-3) provides precise mass and radius estimates for about 10,000 red giants (i.e. corresponding to about two-thirds of their initial sample). These quantities were estimated based on Kepler asteroseismic measurements, \Gaia/DR3 astrometric data \citep{GaiaDR3}, and APOGEE/DR17 spectroscopic parameters \citep{APOGEEDR17}. In the following, we only considered their best-quality parameters,
which belong to their Gold Category. Of their Gold stars, 4,533  are found in our $HQ$ subsample, and 2,019 of them satisfy the adopted differences of 100~K and 0.1~dex in \T\ and \g. 
Their radius and mass are compared in Fig.~\ref{Fig:ValidMR} (top panels). The stellar radii agree very well with a mean relative difference (bias) of only 2\% and an associated standard deviation of 4\%, in the sense of (\gspspec\ - literature)/literature. For the mass comparison, the agreement is also very satisfactory with a mean relative difference of 4\% and a standard deviation of 15\%. Moreover, most of the mass dispersion is clearly produced by the largest \g\ differences. When we only consider the 750 stars with $\Delta$\g<0.05~dex, the dispersion in mass is reduced to 10\%.

For the dwarfs and subgiants, 176 of our $HQ$ stars are also found in the \cite{APOKASCnaines} catalogue of 415 stars.  We recall that this catalogue is based on $Kepler$ asteroseismic and APOGEE/DR13 spectroscopic data. 
Because of the lower statistics, we did not apply any filter on \T\ for the radius comparison, which is again found to be excellent, with a mean relative difference of 2\% and a standard deviation of 4\% (see Fig.~\ref{Fig:ValidMR}, middle panels). For the mass comparison, we selected the 66 stars with an agreement in \g\ better than 0.1~dex. The bias (7\%) and standard deviation (15\%) 
are slightly larger than before, but this is explained by the differences
in the adopted atmospheric parameters. The 
stars that depart most strongly from the 1:1 relation are those with the largest differences in \T\ and/or \g. The larger error bars might also explain part of the dispersion. In any case, this comparison again confirms the high quality of the radius and mass spectroscopy determinations when compared to the APOKASC values.\\

\noindent APO-K2:
This catalogue contains data for about 7,500 RGB and Red Clump
stars, adopting K2-GAP DR3 asteroseismic data \citep{Zinn22}, and, as APOKASC-3, \Gaia/DR3 astrometric and APOGEE/DR17 spectroscopic data. About 2,500 of these giant stars are found in our catalogue, and 1,254 belong to our $HQ$ subsample. The comparison of the two samples is shown in the bottom panels of Fig.~\ref{Fig:ValidMR}, and we also applied the same filtering as for APOKASC-3 on $\Delta$\T\ and $\Delta$\g\ (565 stars were left). As for the other comparison samples, the radii agree very well (bias and dispersion are equal to 2\% and 8\%, respectively). The slightly larger dispersion is simply explained
by the larger number of stars that differ in \T\ by more than 50-100~K. The larger error bars are also associated with the seismic radii of the largest stars. The mass comparison is also affected by the rather large \g\ differences and rather large uncertainties in both samples, although the agreement is still satisfactory (bias of 5\%).

\subsection{Summary of the validations}
\label{Sect:summary}
The above comparisons confirm the high quality of our catalogue values, which rely on the spectroscopic \gspspec\ parameters. 
By selecting homogeneous, large, and high-quality
reference catalogues of stellar radii and masses 
based on completely independent and direct techniques (i.e.
the least model dependent) such as interferometry and
asteroseismology, we found excellent agreement
with the literature. In particular, this agreement 
improved when we considered our high-quality
subsample and when the effective temperatures and 
surface gravities adopted by the different works agreed, leading to even smaller dispersions. 
We emphasise that the agreement for the radii with respect to interferometric and
asteroseismic data is particularly
impressive (almost no biases, and dispersions of a few percent).
This validates the whole method we adopted, that is, not only the radius determination,
but also the derivation of the interstellar reddening, the bolometric corrections, and the luminosities from the \gspspec\ parameters, based on which we estimated $R$ and $M$.
For the mass comparison with asteroseismic data, typical relative differences
are about 5\% with dispersions of 15-20\%. These numbers are improved by a factor of $\sim$2 when we consider the smaller differences in the adopted \T\ and \g, and this confirms the global
very good agreement for such a difficult parameter derived directly from the
spectroscopic data.
Moreover, this section again supports what we showed in Sect.~\ref{Sec:Errors}, that is, the stellar mass is most sensitive to the uncertainty in \g. Therefore, this atmospheric  parameter has not only a significant effect on $M$, but is also the main driver of its uncertainty. Finally, we also point out that some of the comparison sources used in this section have rather high extinctions (e.g. about 10\% of our selected APOKASC-3 stars have $A_G$>0.2, and $\sim$50 of them have $A_G$>0.5). This further reinforces our validation, because even with extinction, accurate stellar radii and masses are recovered.

\section{Example applications of the catalogue}
\label{Sec:Discuss}
In order to illustrate the wide range of possible applications of our catalogue,
we focus this section on different studies starting from exoplanet topics to Galactic ones. We recall that 2D and 3D interstellar dust maps based on our catalogue have been presented by \cite{Marie26}.

\subsection{Radii and masses of exoplanets}
A first example of an application of the \gspspec\ catalogue of stellar radii and masses was presented by \cite{ExoP}, who published a large, homogeneous, and precise catalogue of radii and masses
of 3,556 exoplanets hosted by 2,573 stars. 
These new planet parameters were derived by rescaling the previously published planet parameter values, adopting the stellar radii and masses we estimated here.
The lower uncertainty of the planet radii compared to literature values allowed these authors to improve the characterisation of
the decrease in the number of small planets around 1.8~R$_{\rm Earth}$ \cite[evaporation valley,][]{Fulton17}. Among other results, a dichotomy between dense and inflated planets was also found. Denser planets 
defined by $R_p \le 1.1$~R$_{\rm Jup}$ appeared to be more massive for more metal-rich host stars. Conversely, inflated planets
orbiting more metal-poor stars were found to be more massive. It was proposed that this bimodality might
reveal that the diversity of giant exoplanets might depend on their Galactic birth locus. Dense giant planets were indeed found
to be more numerous than inflated ones for supra-Solar metallicities, as in the central Milky Way regions.
We refer to \cite{ExoP} and forthcoming articles investigating the host star chemical properties for more details on the
\Gaia\ spectroscopic catalogue of exoplanets.

\subsection{Stellar mass distributions}
\begin{figure}[]
    \centering
   \includegraphics[width=0.42\textwidth]{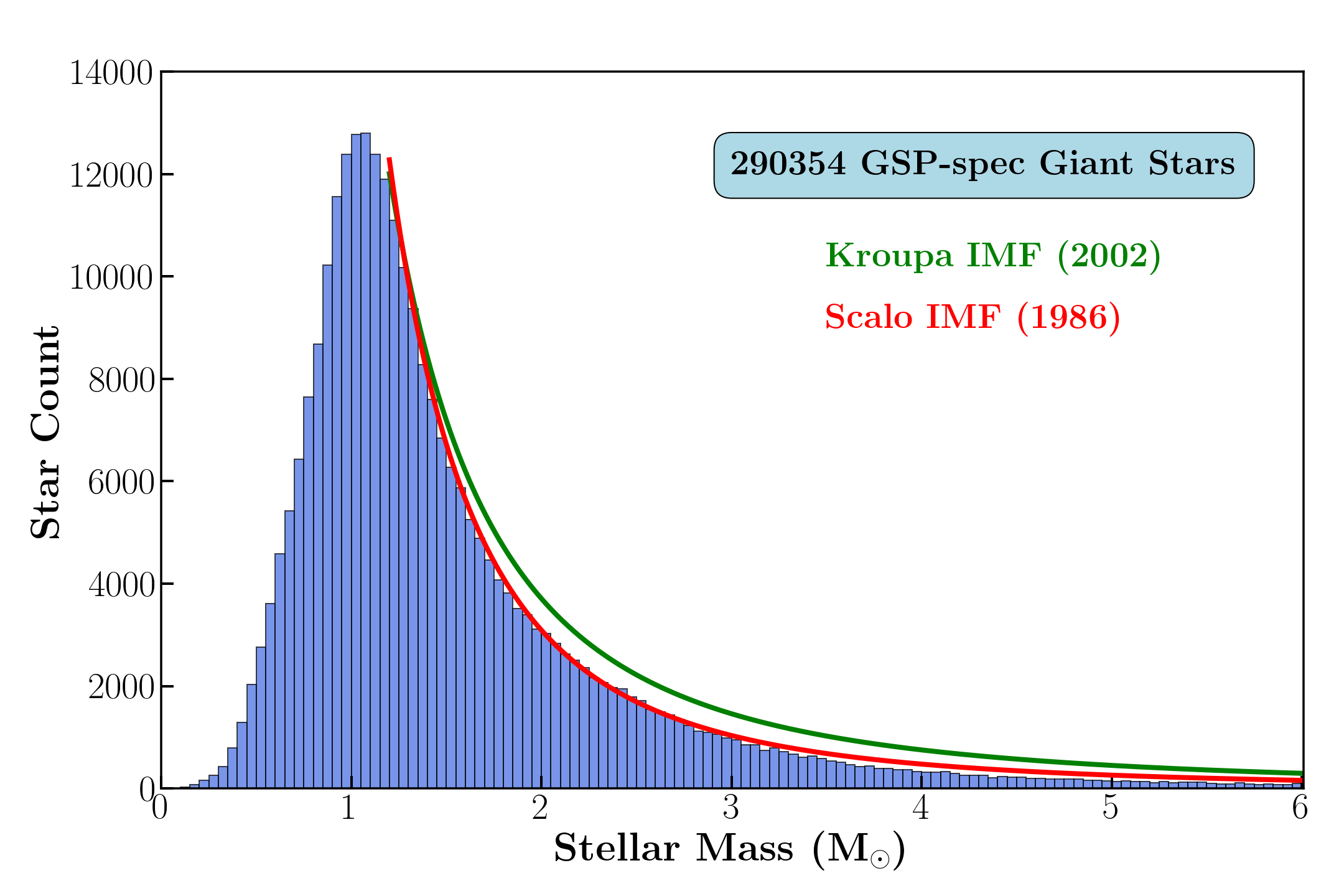} 
    \caption{Present-day mass function of a subsample of bright giant stars (see text for details about the adopted filtering selection) compared to the IMF of Scalo and Kroupa, whose power-law indexes are equal to 2.7 and 2.3, respectively.}
    \label{Fig:IMF}
\end{figure}
\begin{figure}[]
    \centering
   \includegraphics[width=0.42\textwidth]{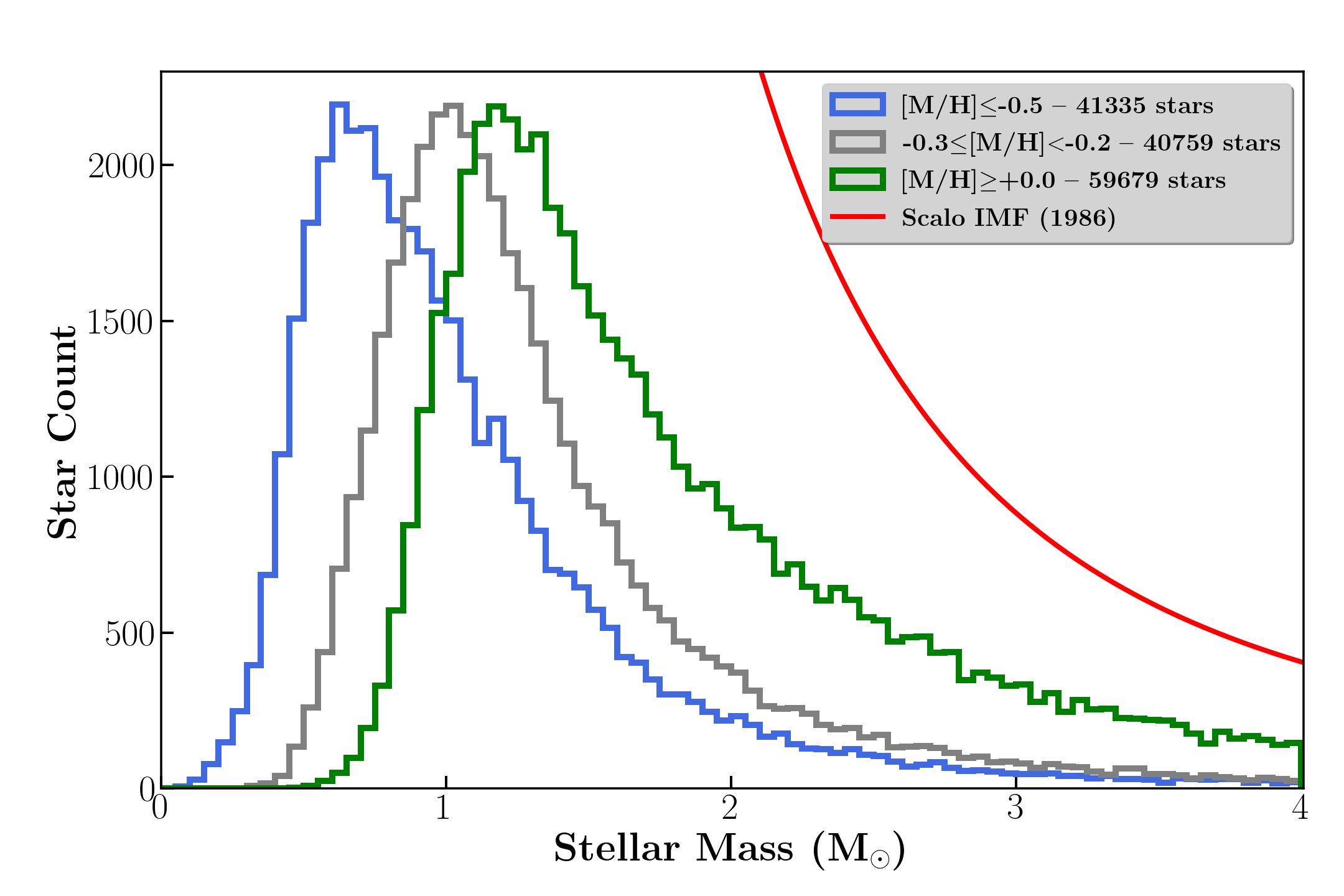} 
    \caption{Same as Fig.~\ref{Fig:IMF}, but for the three metallicity bins
    indicated in the legend. The total number of stars in each histogram is also reported. The Scalo IMF is that of Fig.~\ref{Fig:IMF}.}
    \label{Fig:IMF2}
\end{figure}

In order to explore the stellar mass distribution of our catalogue, which indeed corresponds to the Present-Day Mass Function (PDMF) and to compare
it to theoretical Initial Mass Function (IMF), we first defined
the most complete subsample, that is, a subsample not affected by observational biases. We focused on giant stars to do this because the \gspspec\ catalogue is known to be less complete for dwarfs. Then, following Fig.~2 of \cite{GSPspecDR3}, the rise in \gspspec\ stellar counts with the $G$-magnitude clearly decelerates for stars fainter than $\sim$12.5. 
We therefore show in Fig.~\ref{Fig:IMF} the mass distribution of a subsample of almost 300,000 giant stars found in our catalogue, characterised by $L>200~L_\odot$ and $G<12$. 
Adopting the $HQ$ subsample for our application biases the subsample by favouring high-\SNR\ spectra and leading to a rougher distribution with a number of stars more massive than $\sim2.5~M_\odot$ that is too large (and unrealistic). This figure shows that the PDMF of our giant stars agrees well with the IMF of \citet{Scalo86} over the range $\sim$1.2 - $\sim$3.5~M$_\odot$. Conversely, the \cite{Kroupa} IMF\footnote{The IMF power-law indexes of the Salpeter and Kroupa distributions are too close (2.35 and 2.3) to be distinguishable in this figure.} predicts a number of stars more massive than $\sim 1.5~M_\odot$ that is too large and is not confirmed by our catalogue.
We recall that the Scalo IMF was derived from Galactic field star counts and is therefore a composite IMF of many star-forming regions. It is thus expected that it fits our data well based on stellar counts in the Solar vicinity.

This rather good agreement between our PDMF and the Scalo IMF can be understood
by examining Fig.~\ref{Fig:IMF2}, which shows the PDMF for the same sample of giant stars, but only those belonging to three specific metallicity ranges. These ranges were defined in order to have rather similar number of stars and to span the metal-poor to the metal-rich regime. A rather smooth variation can be seen with a larger number of massive stars with increasing metallicity. but these PDMFs differ from the Scalo IMF, regardless of the normalisation factor. In particular, this figure shows that most stars with supra-Solar metallicities are more massive than the Sun, with several stars having $M>2~M_\odot$. In contrast, almost all the most metal-poor stars (\meta<-0.5~dex) have masses lower than 1.5~M$_\odot$, and their mass distribution is closest to the Scalo IMF.
All of this is quantified in the first column of Tab.~\ref{Tab:Orbits}, which reports the median mass of these three metallicity ranges. The median absolute deviations associated with these medians are similar for the three ranges and equal to $\sim$0.5~M$_\odot$. This increased number of massive stars for higher metallicities can easily be interpreted when we consider that the \gspspec\-sampled volume spans different Galactic populations. We indeed report in Tab.~\ref{Tab:Orbits} some Galactic orbital parameters for the three metallicity bins that were computed by \cite{Pedro23}. In particular, $\sigma$(V$_Z$), the Galactic vertical velocity dispersion can be viewed as an age proxy of the stellar populations. This quantity indeed increases with time \cite[see, for instance,][and references therein]{Hayden18} because of the progressive dynamical 
heating of the disc. $\sigma$(V$_Z$) is larger for the most metal-poor regime composed by lower-mass and thus, older, stars whose orbits have been heated. In contrast, the orbits of the most metal-rich and massive stars, which are younger, did not have time to be significantly perturbed (lower $\sigma$(V$_Z$)). 
All of this is fully consistent with the increasing eccentricities of the Galactic orbits (mean and dispersion, reported in the third column of Tab.~\ref{Tab:Orbits}) with decreasing metallicity. 

We therefore conclude that the agreement between the PDMF and IMF shown in Fig.~\ref{Fig:IMF} is caused by an age effect. The most metal-poor massive stars 
have already ended their life and are thus absent from the blue histogram of Fig.~\ref{Fig:IMF2}, whereas the 
lower-mass stars in the same metallicity bin had enough time to evolve as giants.
Conversely, most of the low-mass metal-rich stars are still on the main-sequence
and are therefore not included in the green histogram. As a consequence, based on the mass distributions alone, we can claim that the most metal-rich bin of Fig.~\ref{Fig:IMF2} is dominated by young stars that mostly belong to the Galactic thin disc, whereas the most metal-poor bin contains older stars of the Galactic thick disc or halo.

\begin{table}[t]
    \caption{Median masses and Galactic orbital properties.}
    \centering
    \begin{tabular}{lccc}
    \hline
     Metallicity range & Mass & $\sigma$(V$_Z$) & Eccentricity\\
     (units) & (M$_\odot$) & (km/s) & \\
    \hline
    \meta $\ge$ 0.0~dex & 1.7 & 22.0 & 0.13$\pm$0.09 \\
    -0.3 $\le$ \meta < -0.2~dex & 1.2 & 27.1 & 0.15$\pm$0.10\\
    \meta $\le$ -0.5~dex & 0.9 & 58.4 & 0.32$\pm$0.25\\
     \hline
    \end{tabular}
    \label{Tab:Orbits}
    \tablefoot{The Galactic orbital properties correspond to the dispersion of the vertical velocity and the mean eccentricity plus its dispersion. The considered stars are the giants belonging to the three metallicity bins shown in Fig.~\ref{Fig:IMF2}.}
\end{table}

\subsection{Galactic halo accreted stars}
\begin{figure}[]
    \centering
   \includegraphics[width=0.45\textwidth]{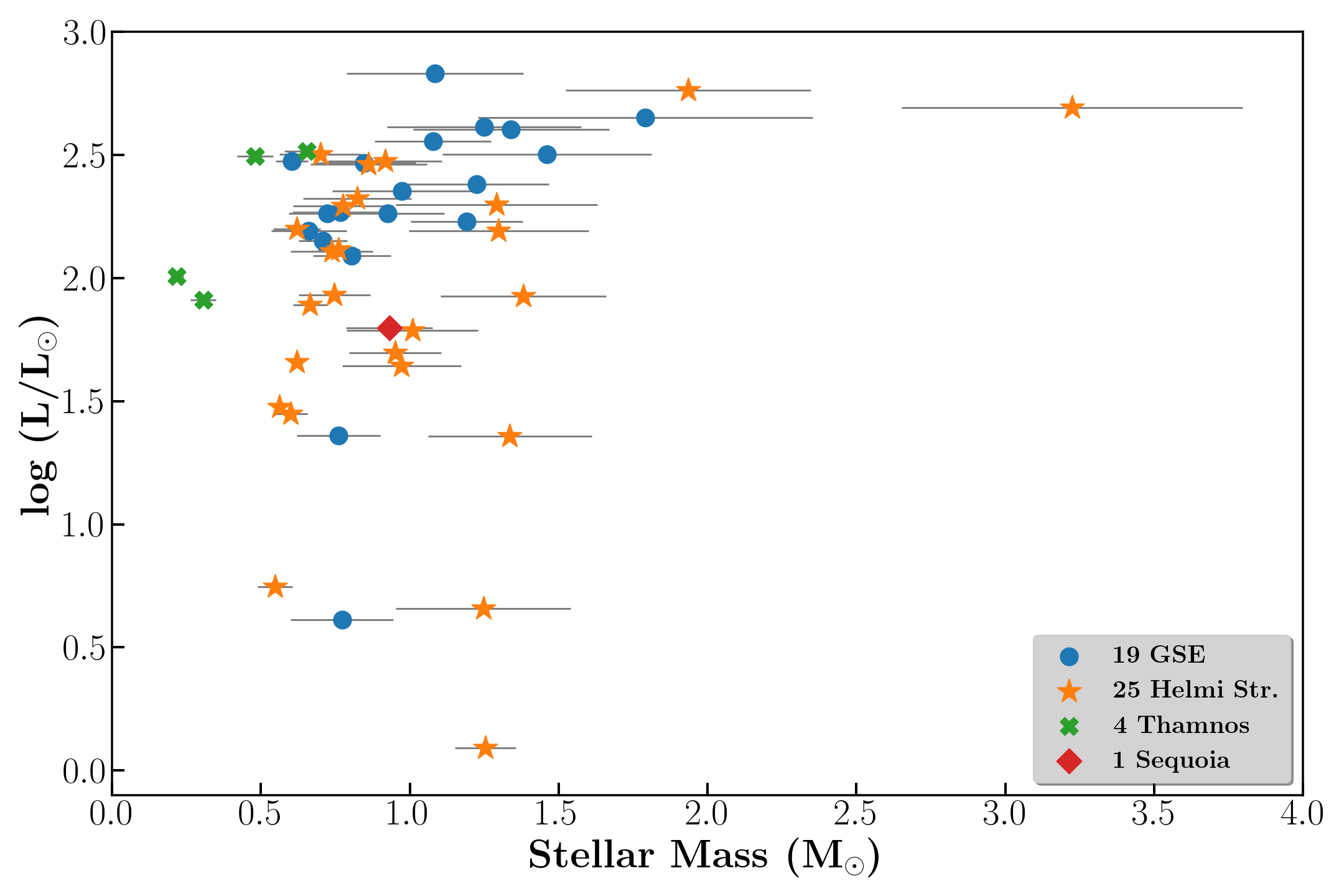} 
    \caption{Mass and luminosity of $HQ$ candidate accreted stars identified through their orbit by \cite{PVP_Ale}. The considered accretion debris systems are indicated in the legend, together with the number of their member stars. The relative uncertainties in luminosity are equal to a few percent and are not visible in the logarithmic scale.}
    \label{Fig:Accreted}
\end{figure}

It is now widely admitted that the Galactic halo is in part composed of several substructures (some of them are called streams)
that are interpreted as signatures of past or ongoing merger events. 
Therefore, some stars belonging to the Galactic halo were formed in external galaxies in ancient epochs. Such merger debris stars can be identified through their specific
Galactic orbits. For instance, they are located in specific regions in a diagram of the total energy ($E$) 
versus $L_Z$ (vertical component of the angular momentum) \citep[see, for instance, Fig.~25-27 in][]{PVP_Ale}. 

We adopted the list of candidate accreted
\gspspec\ stars identified by \cite{PVP_Ale}. They belong to the \Gaia-Enceladus-Sausage accreted galaxy
\cite[GES,][]{Belo18, Myeong18, Helmi18}, the Sequoia \citep{Myeong19}, the Thamnos \citep{Koppelman, Helmi20}, and the Helmi streams \citep{Helmi99}. We cross-matched 
this list with the $HQ$ subsample defined above, including the restriction \g$_{\rm err} <$ 0.1~dex in order to select the most accurate derived masses. We show in Fig.~\ref{Fig:Accreted} the
stellar masses of the accretion debris candidate stars versus their luminosity. 
Most of these stars are high-luminosity stars with masses 
close to or lower than about 1~M$_\odot$ (considering the uncertainties).
These masses are fully consistent with the asteroseismic masses of GES stars derived by \cite{Josefina21}. We also confirmed that these stars are all cool giants that left the main sequence a long time ago. The stellar masses of red giants 
are indeed a good proxy for their age, and typical ages for such post-main-sequence low-mass stars are older than $\sim$10-12~Gy. This is fully compatible with the accretion epochs of the considered mergers. It is indeed thought that the GSE accretion occurred
about 10 billion years ago \citep{Helmi18}, about 5-8~Gyr ago for the Helmi stream \citep{Koppelman}, and between 8-11~Gyr ago for Sequoia; Thamnos is probably older \citep{Dodd}.

There are a few stars in Fig~\ref{Fig:Accreted} that belong to GES or the Helmi stream, however, that are more massive than $\sim$1.5~M$_\odot$ (therefore younger than the other candidates) or too faint to be giants. As already pointed out by \cite{PVP_Ale}, their sample of accreted stars belonging to the Helmi stream
or GES is contaminated by disc stars because these streams lie close to the Galactic disc in the $E - L_Z$ plane.
We therefore confirmed this contamination when we examined the stellar masses and/or luminosities.
Based on these derived quantities, the contamination is stronger for the Helmi stream (it might reach up to $\sim$20~\%) than for GSE, which is closer to $\sim$10~\%. We thus emphasise that a selection of merger debris stars based on purely kinematical criteria should be complemented by an examination of their chemical content \citep[see, for instance, Fig~29 of ][]{PVP_Ale}, but also their mass (as proposed here) and/or age.

\section{Summary}
\label{Sec:Conclu}
We have presented a new catalogue of interstellar colour excesses and extinctions, together with stellar luminosities, radii and masses
for about 4.6 million stars. These data were computed from the \Gaia/\gspspec\ atmospheric parameters (\T, \g, \meta\ , and \AF, derived from the analysis of the RVS spectra), combined with \Gaia/DR3 astrometry and photometry. These parameters are therefore fully compatible and homogeneous with the \Gaia/DR3 spectroscopic data. This procedure allowed us to avoid systematics and biases that might be hidden when data from different (and potentially) heterogeneous catalogues are combined. The uncertainties associated
with all these new parameters were estimated based on 1,000 Monte-Carlo simulations for each star. We also included three
quality flags in this catalogue that should be considered, together with the associated
uncertainties, to define high-quality parameter subsamples.
An example of such a $HQ$ subsample, containing 1.5 million stars, is provided.

The high quality of the catalogue for dwarfs and giants (even when they are highly extincted by the interstellar medium),
was confirmed through different validations based on 
published interstellar maps, 
interferometric radii, and asteroseismic radii and masses. We found that the radius determinations have a similar
precision as interferometric and asteroseismic determinations when similar effective temperatures were considered by us and the
asteroseismic works. The relative mean differences between our mass determinations and the Kepler determinations are as small as a few percent,
associated with a dispersion of 10-20\%, when the agreement in surface
gravity is good. This new catalogue, its large statistics, homogeneity, and high quality allowed us to explore different astrophysical problematics. Some example applications were described from exoplanets \citep{ExoP} to properties of the Galactic halo. For instance, we found that (i) the Present-Day Mass Function of the brightest giant stars is compatible with the Kroupa IMF when any metallicities are considered because of the different evolutionary timescales of low- and high-mass stars found in the different
Galactic populations, (ii) the main Galactic populations can be identified directly based on their mass distribution, and (iii) the estimated low masses for the accreted halo stars are compatible with the epoch of their accretion into the Milky Way.

We finally emphasise that part of the methods we presented
were adopted to calibrate the \Gaia/DR4 \gspspec\ data that will be published in December 2026
(Palicio et al., in preparation). Moreover, a new version of this catalogue will be published after the \Gaia/DR4, based on the new atmospheric parameter values derived from the RVS spectra. Its precision will be improved by the new implementations in the \gspspec\ pipeline. Improvements in the parameter accuracy are also expected because the uncertainties associated with the DR4 stellar parameters are lower since the S/N of the RVS spectra increases from DR3 to DR4. This catalogue size will also be extended by about a factor of 5-6 (i.e. a few tens of million stars) because more spectra are parametrised by \gspspec\ for the fourth \Gaia\ data release.

\begin{acknowledgements}
This work has made use of data from the European Space Agency (ESA)
mission \Gaia\ (https://www.cosmos.esa.int/gaia), processed by the \Gaia\ Data Processing and Analysis Consortium (DPAC, https://www.cosmos.esa.int/web/gaia/dpac/consortium). Funding for the DPAC has been provided by national institutions, in particular the institutions participating in the Gaia Multilateral Agreement.\\

We acknowledge funding from the European
Union’s Horizon 2020 research and innovation program under SPACE-H2020
grant agreement number 101004214 (EXPLORE project).\\

This work has made use of the IPython package \citep{ipython}, NumPy \citep{NumPy}, Matplotlib \citep{Matplotlib}, Pandas, TOPCAT \citep{Topcat} and the SIMBAD database,
operated at CDS, Strasbourg, France \citep{Simbad}. We also thank L. Casagrande for helpful discussions and the anonymous referee for his sensible comments. E.S thanks I.N.A.F. for the 1.05.24.07.02 Mini Grant – LEGARE “Linking the chemical Evolution of Galactic discs AcRoss diversE scales: from the
thin disc to the nuclear stellar disc” (PI E. Spitoni)
\end{acknowledgements}

\bibliographystyle{aa} 
\bibliography{ref}

\begin{thebibliography}{49}
\expandafter\ifx\csname natexlab\endcsname\relax\def\natexlab#1{#1}\fi

\bibitem[{{Abdurro'uf} {et~al.}(2022){Abdurro'uf}, {Accetta}, {Aerts}, {Silva Aguirre}, {Ahumada}, {Ajgaonkar}, {Filiz Ak}, {Alam}, {Allende Prieto}, {Almeida}, {Anders}, {Anderson}, {Andrews}, {Anguiano}, {Aquino-Ort{\'\i}z}, {Arag{\'o}n-Salamanca}, {Argudo-Fern{\'a}ndez}, {Ata}, {Aubert}, {Avila-Reese}, {Badenes}, {Barb{\'a}}, {Barger}, {Barrera-Ballesteros}, {Beaton}, {Beers}, {Belfiore}, {Bender}, {Bernardi}, {Bershady}, {Beutler}, {Bidin}, {Bird}, {Bizyaev}, {Blanc}, {Blanton}, {Boardman}, {Bolton}, {Boquien}, {Borissova}, {Bovy}, {Brandt}, {Brown}, {Brownstein}, {Brusa}, {Buchner}, {Bundy}, {Burchett}, {Bureau}, {Burgasser}, {Cabang}, {Campbell}, {Cappellari}, {Carlberg}, {Wanderley}, {Carrera}, {Cash}, {Chen}, {Chen}, {Cherinka}, {Chiappini}, {Choi}, {Chojnowski}, {Chung}, {Clerc}, {Cohen}, {Comerford}, {Comparat}, {da Costa}, {Covey}, {Crane}, {Cruz-Gonzalez}, {Culhane}, {Cunha}, {Dai}, {Damke}, {Darling}, {Davidson}, {Davies}, {Dawson}, {De Lee}, {Diamond-Stanic}, {Cano-D{\'\i}az}, {S{\'a}nchez},
  {Donor}, {Duckworth}, {Dwelly}, {Eisenstein}, {Elsworth}, {Emsellem}, {Eracleous}, {Escoffier}, {Fan}, {Farr}, {Feng}, {Fern{\'a}ndez-Trincado}, {Feuillet}, {Filipp}, {Fillingham}, {Frinchaboy}, {Fromenteau}, {Galbany}, {Garc{\'\i}a}, {Garc{\'\i}a-Hern{\'a}ndez}, {Ge}, {Geisler}, {Gelfand}, {G{\'e}ron}, {Gibson}, {Goddy}, {Godoy-Rivera}, {Grabowski}, {Green}, {Greener}, {Grier}, {Griffith}, {Guo}, {Guy}, {Hadjara}, {Harding}, {Hasselquist}, {Hayes}, {Hearty}, {Hern{\'a}ndez}, {Hill}, {Hogg}, {Holtzman}, {Horta}, {Hsieh}, {Hsu}, {Hsu}, {Huber}, {Huertas-Company}, {Hutchinson}, {Hwang}, {Ibarra-Medel}, {Chitham}, {Ilha}, {Imig}, {Jaekle}, {Jayasinghe}, {Ji}, {Johnson}, {Jones}, {J{\"o}nsson}, {Katkov}, {Khalatyan}, {Kinemuchi}, {Kisku}, {Knapen}, {Kneib}, {Kollmeier}, {Kong}, {Kounkel}, {Kreckel}, {Krishnarao}, {Lacerna}, {Lane}, {Langgin}, {Lavender}, {Law}, {Lazarz}, {Leung}, {Leung}, {Lewis}, {Li}, {Li}, {Lian}, {Liang}, {Lin}, {Lin}, {Lin}, {Lintott}, {Long}, {Longa-Pe{\~n}a}, {L{\'o}pez-Cob{\'a}}, {Lu},
  {Lundgren}, {Luo}, {Mackereth}, {de la Macorra}, {Mahadevan}, {Majewski}, {Manchado}, {Mandeville}, {Maraston}, {Margalef-Bentabol}, {Masseron}, {Masters}, {Mathur}, {McDermid}, {Mckay}, {Merloni}, {Merrifield}, {Meszaros}, {Miglio}, {Di Mille}, {Minniti}, {Minsley}, {Monachesi}, {Moon}, {Mosser}, {Mulchaey}, {Muna}, {Mu{\~n}oz}, {Myers}, {Myers}, {Nadathur}, {Nair}, {Nandra}, {Neumann}, {Newman}, {Nidever}, {Nikakhtar}, {Nitschelm}, {O'Connell}, {Garma-Oehmichen}, {Luan Souza de Oliveira}, {Olney}, {Oravetz}, {Ortigoza-Urdaneta}, {Osorio}, {Otter}, {Pace}, {Padilla}, {Pan}, {Pan}, {Parikh}, {Parker}, {Peirani}, {Pe{\~n}a Ram{\'\i}rez}, {Penny}, {Percival}, {Perez-Fournon}, {Pinsonneault}, {Poidevin}, {Poovelil}, {Price-Whelan}, {B{\'a}rbara de Andrade Queiroz}, {Raddick}, {Ray}, {Rembold}, {Riddle}, {Riffel}, {Riffel}, {Rix}, {Robin}, {Rodr{\'\i}guez-Puebla}, {Roman-Lopes}, {Rom{\'a}n-Z{\'u}{\~n}iga}, {Rose}, {Ross}, {Rossi}, {Rubin}, {Salvato}, {S{\'a}nchez}, {S{\'a}nchez-Gallego}, {Sanderson}, {Santana
  Rojas}, {Sarceno}, {Sarmiento}, {Sayres}, {Sazonova}, {Schaefer}, {Schiavon}, {Schlegel}, {Schneider}, {Schultheis}, {Schwope}, {Serenelli}, {Serna}, {Shao}, {Shapiro}, {Sharma}, {Shen}, {Shetrone}, {Shu}, {Simon}, {Skrutskie}, {Smethurst}, {Smith}, {Sobeck}, {Spoo}, {Sprague}, {Stark}, {Stassun}, {Steinmetz}, {Stello}, {Stone-Martinez}, {Storchi-Bergmann}, {Stringfellow}, {Stutz}, {Su}, {Taghizadeh-Popp}, {Talbot}, {Tayar}, {Telles}, {Teske}, {Thakar}, {Theissen}, {Tkachenko}, {Thomas}, {Tojeiro}, {Hernandez Toledo}, {Troup}, {Trump}, {Trussler}, {Turner}, {Tuttle}, {Unda-Sanzana}, {V{\'a}zquez-Mata}, {Valentini}, {Valenzuela}, {Vargas-Gonz{\'a}lez}, {Vargas-Maga{\~n}a}, {Alfaro}, {Villanova}, {Vincenzo}, {Wake}, {Warfield}, {Washington}, {Weaver}, {Weijmans}, {Weinberg}, {Weiss}, {Westfall}, {Wild}, {Wilde}, {Wilson}, {Wilson}, {Wilson}, {Wolf}, {Wood-Vasey}, {Yan}, {Zamora}, {Zasowski}, {Zhang}, {Zhao}, {Zheng}, {Zheng}, \& {Zhu}}]{APOGEEDR17}
{Abdurro'uf}, {Accetta}, K., {Aerts}, C., {et~al.} 2022, \apjs, 259, 35

\bibitem[{{Bailer-Jones} {et~al.}(2021){Bailer-Jones}, {Rybizki}, {Fouesneau}, {Demleitner}, \& {Andrae}}]{Coryn21}
{Bailer-Jones}, C.~A.~L., {Rybizki}, J., {Fouesneau}, M., {Demleitner}, M., \& {Andrae}, R. 2021, \aj, 161, 147

\bibitem[{{Barbillon} {et~al.}(2025){Barbillon}, {Recio-Blanco}, {de Laverny}, \& {Palicio}}]{Marie26}
{Barbillon}, M., {Recio-Blanco}, A., {de Laverny}, P., \& {Palicio}, P.~A. 2025, arXiv e-prints, arXiv:2511.12156

\bibitem[{{Belokurov} {et~al.}(2018){Belokurov}, {Erkal}, {Evans}, {Koposov}, \& {Deason}}]{Belo18}
{Belokurov}, V., {Erkal}, D., {Evans}, N.~W., {Koposov}, S.~E., \& {Deason}, A.~J. 2018, \mnras, 478, 611

\bibitem[{{Boltzmann}(1884)}]{Boltzmann1884}
{Boltzmann}, L. 1884, Annalen der Physik, 258, 291

\bibitem[{{Borucki} {et~al.}(2008){Borucki}, {Koch}, {Basri}, {Batalha}, {Brown}, {Caldwell}, {Christensen-Dalsgaard}, {Cochran}, {Dunham}, {Gautier}, {Geary}, {Gilliland}, {Jenkins}, {Kondo}, {Latham}, {Lissauer}, \& {Monet}}]{Kepler}
{Borucki}, W., {Koch}, D., {Basri}, G., {et~al.} 2008, in IAU Symposium, Vol. 249, Exoplanets: Detection, Formation and Dynamics, ed. Y.-S. {Sun}, S.~{Ferraz-Mello}, \& J.-L. {Zhou}, 17--24

\bibitem[{{Casagrande} {et~al.}(2021){Casagrande}, {Lin}, {Rains}, {Liu}, {Buder}, {Horner}, {Asplund}, {Lewis}, {Martell}, {Nordlander}, {Stello}, {Ting}, {Wittenmyer}, {Bland-Hawthorn}, {Casey}, {De Silva}, {D'Orazi}, {Freeman}, {Hayden}, {Kos}, {Lind}, {Schlesinger}, {Sharma}, {Simpson}, {Zucker}, \& {Zwitter}}]{Luca21}
{Casagrande}, L., {Lin}, J., {Rains}, A.~D., {et~al.} 2021, \mnras, 507, 2684

\bibitem[{{Casagrande} \& {VandenBerg}(2018)}]{Luca18}
{Casagrande}, L. \& {VandenBerg}, D.~A. 2018, \mnras, 479, L102

\bibitem[{{Creevey} {et~al.}(2023){Creevey}, {Sordo}, {Pailler}, {Fr{\'e}mat}, {Heiter}, {Th{\'e}venin}, {Andrae}, {Fouesneau}, {Lobel}, {Bailer-Jones}, {Garabato}, {Bellas-Velidis}, {Brugaletta}, {Lorca}, {Ordenovic}, {Palicio}, {Sarro}, {Delchambre}, {Drimmel}, {Rybizki}, {Torralba Elipe}, {Korn}, {Recio-Blanco}, {Schultheis}, {De Angeli}, {Montegriffo}, {Abreu Aramburu}, {Accart}, {{\'A}lvarez}, {Bakker}, {Brouillet}, {Burlacu}, {Carballo}, {Casamiquela}, {Chiavassa}, {Contursi}, {Cooper}, {Dafonte}, {Dapergolas}, {de Laverny}, {Dharmawardena}, {Edvardsson}, {Le Fustec}, {Garc{\'\i}a-Lario}, {Garc{\'\i}a-Torres}, {Gomez}, {Gonz{\'a}lez-Santamar{\'\i}a}, {Hatzidimitriou}, {Jean-Antoine Piccolo}, {Kontiza}, {Kordopatis}, {Lanzafame}, {Lebreton}, {Licata}, {Lindstr{\o}m}, {Livanou}, {Magdaleno Romeo}, {Manteiga}, {Marocco}, {Marshall}, {Mary}, {Nicolas}, {Pallas-Quintela}, {Panem}, {Pichon}, {Poggio}, {Riclet}, {Robin}, {Santove{\~n}a}, {Silvelo}, {Slezak}, {Smart}, {Soubiran}, {S{\"u}veges}, {Ulla},
  {Utrilla}, {Vallenari}, {Zhao}, {Zorec}, {Barrado}, {Bijaoui}, {Bouret}, {Blomme}, {Brott}, {Cassisi}, {Kochukhov}, {Martayan}, {Shulyak}, \& {Silvester}}]{Orlagh23}
{Creevey}, O.~L., {Sordo}, R., {Pailler}, F., {et~al.} 2023, \aap, 674, A26

\bibitem[{{Cropper} {et~al.}(2018){Cropper}, {Katz}, {Sartoretti}, {Prusti}, {de Bruijne}, {Chassat}, {Charvet}, {Boyadjian}, {Perryman}, {Sarri}, {Gare}, {Erdmann}, {Munari}, {Zwitter}, {Wilkinson}, {Arenou}, {Vallenari}, {G{\'o}mez}, {Panuzzo}, {Seabroke}, {Allende Prieto}, {Benson}, {Marchal}, {Huckle}, {Smith}, {Dolding}, {Jan{\ss}en}, {Viala}, {Blomme}, {Baker}, {Boudreault}, {Crifo}, {Soubiran}, {Fr{\'e}mat}, {Jasniewicz}, {Guerrier}, {Guy}, {Turon}, {Jean-Antoine-Piccolo}, {Th{\'e}venin}, {David}, {Gosset}, \& {Damerdji}}]{RVS}
{Cropper}, M., {Katz}, D., {Sartoretti}, P., {et~al.} 2018, \aap, 616, A5

\bibitem[{{de Laverny} {et~al.}(2025){de Laverny}, {Ligi}, {Crida}, {Recio-Blanco}, \& {Palicio}}]{ExoP}
{de Laverny}, P., {Ligi}, R., {Crida}, A., {Recio-Blanco}, A., \& {Palicio}, P.~A. 2025, \aap, 699, A100

\bibitem[{{de Laverny} {et~al.}(2024){de Laverny}, {Recio-Blanco}, {Aerts}, \& {Palicio}}]{GamDor}
{de Laverny}, P., {Recio-Blanco}, A., {Aerts}, C., \& {Palicio}, P.~A. 2024, \aap, 691, A182

\bibitem[{{Dodd} {et~al.}(2025){Dodd}, {Ruiz-Lara}, {Helmi}, {Gallart}, {Callingham}, {Cassisi}, {Fern{\'a}ndez-Alvar}, \& {Surot}}]{Dodd}
{Dodd}, E., {Ruiz-Lara}, T., {Helmi}, A., {et~al.} 2025, \aap, 698, A277

\bibitem[{{Duvert}(2016)}]{DuvertJMMC}
{Duvert}, G. 2016, {VizieR Online Data Catalog: JMDC : JMMC Measured Stellar Diameters Catalogue (Duvert, 2016)}, VizieR On-line Data Catalog: II/345. Originally published in: JMMC center (2016)

\bibitem[{{Fulton} {et~al.}(2017){Fulton}, {Petigura}, {Howard}, {Isaacson}, {Marcy}, {Cargile}, {Hebb}, {Weiss}, {Johnson}, {Morton}, {Sinukoff}, {Crossfield}, \& {Hirsch}}]{Fulton17}
{Fulton}, B.~J., {Petigura}, E.~A., {Howard}, A.~W., {et~al.} 2017, \aj, 154, 109

\bibitem[{{Gaia Collaboration} {et~al.}(2023{\natexlab{a}}){Gaia Collaboration}, {Recio-Blanco}, {Kordopatis}, {de Laverny}, {Palicio}, {Spagna}, {Spina}, {Katz}, {Re Fiorentin}, {Poggio}, {McMillan}, {Vallenari}, {Lattanzi}, {Seabroke}, {Casamiquela}, {Bragaglia}, {Antoja}, {Bailer-Jones}, {Schultheis}, {Andrae}, {Fouesneau}, {Cropper}, {Cantat-Gaudin}, {Bijaoui}, {Heiter}, {Brown}, {Prusti}, {de Bruijne}, {Arenou}, {Babusiaux}, {Biermann}, {Creevey}, {Ducourant}, {Evans}, {Eyer}, {Guerra}, {Hutton}, {Jordi}, {Klioner}, {Lammers}, {Lindegren}, {Luri}, {Mignard}, {Panem}, {Pourbaix}, {Randich}, {Sartoretti}, {Soubiran}, {Tanga}, {Walton}, {Bastian}, {Drimmel}, {Jansen}, {van Leeuwen}, {Bakker}, {Cacciari}, {Casta{\~n}eda}, {De Angeli}, {Fabricius}, {Fr{\'e}mat}, {Galluccio}, {Guerrier}, {Masana}, {Messineo}, {Mowlavi}, {Nicolas}, {Nienartowicz}, {Pailler}, {Panuzzo}, {Riclet}, {Roux}, {Sordo}, {Th{\'e}venin}, {Gracia-Abril}, {Portell}, {Teyssier}, {Altmann}, {Audard}, {Bellas-Velidis}, {Benson}, {Berthier},
  {Blomme}, {Burgess}, {Busonero}, {Busso}, {C{\'a}novas}, {Carry}, {Cellino}, {Cheek}, {Clementini}, {Damerdji}, {Davidson}, {de Teodoro}, {Nu{\~n}ez Campos}, {Delchambre}, {Dell'Oro}, {Esquej}, {Fern{\'a}ndez-Hern{\'a}ndez}, {Fraile}, {Garabato}, {Garc{\'\i}a-Lario}, {Gosset}, {Haigron}, {Halbwachs}, {Hambly}, {Harrison}, {Hern{\'a}ndez}, {Hestroffer}, {Hodgkin}, {Holl}, {Jan{\ss}en}, {Jevardat de Fombelle}, {Jordan}, {Krone-Martins}, {Lanzafame}, {L{\"o}ffler}, {Marchal}, {Marrese}, {Moitinho}, {Muinonen}, {Osborne}, {Pancino}, {Pauwels}, {Reyl{\'e}}, {Riello}, {Rimoldini}, {Roegiers}, {Rybizki}, {Sarro}, {Siopis}, {Smith}, {Sozzetti}, {Utrilla}, {van Leeuwen}, {Abbas}, {{\'A}brah{\'a}m}, {Abreu Aramburu}, {Aerts}, {Aguado}, {Ajaj}, {Aldea-Montero}, {Altavilla}, {{\'A}lvarez}, {Alves}, {Anders}, {Anderson}, {Anglada Varela}, {Baines}, {Baker}, {Balaguer-N{\'u}{\~n}ez}, {Balbinot}, {Balog}, {Barache}, {Barbato}, {Barros}, {Barstow}, {Bartolom{\'e}}, {Bassilana}, {Bauchet}, {Becciani}, {Bellazzini},
  {Berihuete}, {Bernet}, {Bertone}, {Bianchi}, {Binnenfeld}, {Blanco-Cuaresma}, {Boch}, {Bombrun}, {Bossini}, {Bouquillon}, {Bramante}, {Breedt}, {Bressan}, {Brouillet}, {Brugaletta}, {Bucciarelli}, {Burlacu}, {Butkevich}, {Buzzi}, {Caffau}, {Cancelliere}, {Carballo}, {Carlucci}, {Carnerero}, {Carrasco}, {Castellani}, {Castro-Ginard}, {Chaoul}, {Charlot}, {Chemin}, {Chiaramida}, {Chiavassa}, {Chornay}, {Comoretto}, {Contursi}, {Cooper}, {Cornez}, {Cowell}, \& {Crifo}}]{PVP_Ale}
{Gaia Collaboration}, {Recio-Blanco}, A., {Kordopatis}, G., {et~al.} 2023{\natexlab{a}}, \aap, 674, A38

\bibitem[{{Gaia Collaboration} {et~al.}(2023{\natexlab{b}}){Gaia Collaboration}, {Vallenari}, {Brown}, {Prusti}, {de Bruijne}, {Arenou}, {Babusiaux}, {Biermann}, {Creevey}, {Ducourant}, {Evans}, {Eyer}, {Guerra}, {Hutton}, {Jordi}, {Klioner}, {Lammers}, {Lindegren}, {Luri}, {Mignard}, {Panem}, {Pourbaix}, {Randich}, {Sartoretti}, {Soubiran}, {Tanga}, {Walton}, {Bailer-Jones}, {Bastian}, {Drimmel}, {Jansen}, {Katz}, {Lattanzi}, {van Leeuwen}, {Bakker}, {Cacciari}, {Casta{\~n}eda}, {De Angeli}, {Fabricius}, {Fouesneau}, {Fr{\'e}mat}, {Galluccio}, {Guerrier}, {Heiter}, {Masana}, {Messineo}, {Mowlavi}, {Nicolas}, {Nienartowicz}, {Pailler}, {Panuzzo}, {Riclet}, {Roux}, {Seabroke}, {Sordo}, {Th{\'e}venin}, {Gracia-Abril}, {Portell}, {Teyssier}, {Altmann}, {Andrae}, {Audard}, {Bellas-Velidis}, {Benson}, {Berthier}, {Blomme}, {Burgess}, {Busonero}, {Busso}, {C{\'a}novas}, {Carry}, {Cellino}, {Cheek}, {Clementini}, {Damerdji}, {Davidson}, {de Teodoro}, {Nu{\~n}ez Campos}, {Delchambre}, {Dell'Oro}, {Esquej},
  {Fern{\'a}ndez-Hern{\'a}ndez}, {Fraile}, {Garabato}, {Garc{\'\i}a-Lario}, {Gosset}, {Haigron}, {Halbwachs}, {Hambly}, {Harrison}, {Hern{\'a}ndez}, {Hestroffer}, {Hodgkin}, {Holl}, {Jan{\ss}en}, {Jevardat de Fombelle}, {Jordan}, {Krone-Martins}, {Lanzafame}, {L{\"o}ffler}, {Marchal}, {Marrese}, {Moitinho}, {Muinonen}, {Osborne}, {Pancino}, {Pauwels}, {Recio-Blanco}, {Reyl{\'e}}, {Riello}, {Rimoldini}, {Roegiers}, {Rybizki}, {Sarro}, {Siopis}, {Smith}, {Sozzetti}, {Utrilla}, {van Leeuwen}, {Abbas}, {{\'A}brah{\'a}m}, {Abreu Aramburu}, {Aerts}, {Aguado}, {Ajaj}, {Aldea-Montero}, {Altavilla}, {{\'A}lvarez}, {Alves}, {Anders}, {Anderson}, {Anglada Varela}, {Antoja}, {Baines}, {Baker}, {Balaguer-N{\'u}{\~n}ez}, {Balbinot}, {Balog}, {Barache}, {Barbato}, {Barros}, {Barstow}, {Bartolom{\'e}}, {Bassilana}, {Bauchet}, {Becciani}, {Bellazzini}, {Berihuete}, {Bernet}, {Bertone}, {Bianchi}, {Binnenfeld}, {Blanco-Cuaresma}, {Blazere}, {Boch}, {Bombrun}, {Bossini}, {Bouquillon}, {Bragaglia}, {Bramante}, {Breedt},
  {Bressan}, {Brouillet}, {Brugaletta}, {Bucciarelli}, {Burlacu}, {Butkevich}, {Buzzi}, {Caffau}, {Cancelliere}, {Cantat-Gaudin}, {Carballo}, {Carlucci}, {Carnerero}, {Carrasco}, {Casamiquela}, {Castellani}, {Castro-Ginard}, {Chaoul}, {Charlot}, {Chemin}, {Chiaramida}, {Chiavassa}, {Chornay}, {Comoretto}, {Contursi}, {Cooper}, {Cornez}, {Cowell}, {Crifo}, {Cropper}, {Crosta}, {Crowley}, {Dafonte}, {Dapergolas}, {David}, {David}, {de Laverny}, {De Luise}, \& {De March}}]{GaiaDR3}
{Gaia Collaboration}, {Vallenari}, A., {Brown}, A.~G.~A., {et~al.} 2023{\natexlab{b}}, \aap, 674, A1

\bibitem[{Harris {et~al.}(2020)Harris, Millman, van~der Walt, Gommers, Virtanen, Cournapeau, Wieser, Taylor, Berg, Smith, Kern, Picus, Hoyer, van Kerkwijk, Brett, Haldane, del R{'{\i}}o, Wiebe, Peterson, G{'{e}}rard-Marchant, Sheppard, Reddy, Weckesser, Abbasi, Gohlke, \& Oliphant}]{NumPy}
Harris, C.~R., Millman, K.~J., van~der Walt, S.~J., {et~al.} 2020, Nature, 585, 357

\bibitem[{{Hayden} {et~al.}(2018){Hayden}, {Recio-Blanco}, {de Laverny}, {Mikolaitis}, {Guiglion}, {Hill}, {Gilmore}, {Randich}, {Bayo}, {Bensby}, {Bergemann}, {Bragaglia}, {Casey}, {Costado}, {Feltzing}, {Franciosini}, {Hourihane}, {Jofre}, {Koposov}, {Kordopatis}, {Lanzafame}, {Lardo}, {Lewis}, {Lind}, {Magrini}, {Monaco}, {Morbidelli}, {Pancino}, {Sacco}, {Stonkute}, {Worley}, \& {Zwitter}}]{Hayden18}
{Hayden}, M.~R., {Recio-Blanco}, A., {de Laverny}, P., {et~al.} 2018, \aap, 609, A79

\bibitem[{{Helmi}(2020)}]{Helmi20}
{Helmi}, A. 2020, \araa, 58, 205

\bibitem[{{Helmi} {et~al.}(2018){Helmi}, {Babusiaux}, {Koppelman}, {Massari}, {Veljanoski}, \& {Brown}}]{Helmi18}
{Helmi}, A., {Babusiaux}, C., {Koppelman}, H.~H., {et~al.} 2018, \nat, 563, 85

\bibitem[{{Helmi} {et~al.}(1999){Helmi}, {White}, {de Zeeuw}, \& {Zhao}}]{Helmi99}
{Helmi}, A., {White}, S. D.~M., {de Zeeuw}, P.~T., \& {Zhao}, H. 1999, \nat, 402, 53

\bibitem[{Hunter(2007)}]{Matplotlib}
Hunter, J.~D. 2007, Computing In Science \& Engineering, 9, 90

\bibitem[{{Koppelman} {et~al.}(2019){Koppelman}, {Helmi}, {Massari}, {Price-Whelan}, \& {Starkenburg}}]{Koppelman}
{Koppelman}, H.~H., {Helmi}, A., {Massari}, D., {Price-Whelan}, A.~M., \& {Starkenburg}, T.~K. 2019, \aap, 631, L9

\bibitem[{{Kroupa}(2002)}]{Kroupa}
{Kroupa}, P. 2002, Science, 295, 82

\bibitem[{{Miglio} {et~al.}(2013){Miglio}, {Chiappini}, {Morel}, {Barbieri}, {Chaplin}, {Girardi}, {Montalb{\'a}n}, {Valentini}, {Mosser}, {Baudin}, {Casagrande}, {Fossati}, {Silva Aguirre}, \& {Baglin}}]{Miglio13}
{Miglio}, A., {Chiappini}, C., {Morel}, T., {et~al.} 2013, \mnras, 429, 423

\bibitem[{{Montalb{\'a}n} {et~al.}(2021){Montalb{\'a}n}, {Mackereth}, {Miglio}, {Vincenzo}, {Chiappini}, {Buldgen}, {Mosser}, {Noels}, {Scuflaire}, {Vrard}, {Willett}, {Davies}, {Hall}, {Nielsen}, {Khan}, {Rendle}, {van Rossem}, {Ferguson}, \& {Chaplin}}]{Josefina21}
{Montalb{\'a}n}, J., {Mackereth}, J.~T., {Miglio}, A., {et~al.} 2021, Nature Astronomy, 5, 640

\bibitem[{{Myeong} {et~al.}(2018){Myeong}, {Evans}, {Belokurov}, {Sanders}, \& {Koposov}}]{Myeong18}
{Myeong}, G.~C., {Evans}, N.~W., {Belokurov}, V., {Sanders}, J.~L., \& {Koposov}, S.~E. 2018, \mnras, 478, 5449

\bibitem[{{Myeong} {et~al.}(2019){Myeong}, {Vasiliev}, {Iorio}, {Evans}, \& {Belokurov}}]{Myeong19}
{Myeong}, G.~C., {Vasiliev}, E., {Iorio}, G., {Evans}, N.~W., \& {Belokurov}, V. 2019, \mnras, 488, 1235

\bibitem[{{Navarrete} {et~al.}(2025){Navarrete}, {Recio-Blanco}, {de Laverny}, \& {Escorza}}]{Camila}
{Navarrete}, C., {Recio-Blanco}, A., {de Laverny}, P., \& {Escorza}, A. 2025, \aap, 696, A82

\bibitem[{{Palicio} {et~al.}(2023){Palicio}, {Recio-Blanco}, {Poggio}, {Antoja}, {McMillan}, \& {Spitoni}}]{Pedro23}
{Palicio}, P.~A., {Recio-Blanco}, A., {Poggio}, E., {et~al.} 2023, \aap, 670, L7

\bibitem[{P\'erez \& Granger(2007)}]{ipython}
P\'erez, F. \& Granger, B.~E. 2007, Computing in Science and Engineering, 9, 21

\bibitem[{{Pinsonneault} {et~al.}(2018){Pinsonneault}, {Elsworth}, {Tayar}, {Serenelli}, {Stello}, {Zinn}, {Mathur}, {Garc{\'\i}a}, {Johnson}, {Hekker}, {Huber}, {Kallinger}, {M{\'e}sz{\'a}ros}, {Mosser}, {Stassun}, {Girardi}, {Rodrigues}, {Silva Aguirre}, {An}, {Basu}, {Chaplin}, {Corsaro}, {Cunha}, {Garc{\'\i}a-Hern{\'a}ndez}, {Holtzman}, {J{\"o}nsson}, {Shetrone}, {Smith}, {Sobeck}, {Stringfellow}, {Zamora}, {Beers}, {Fern{\'a}ndez-Trincado}, {Frinchaboy}, {Hearty}, \& {Nitschelm}}]{APOKASC2}
{Pinsonneault}, M.~H., {Elsworth}, Y.~P., {Tayar}, J., {et~al.} 2018, \apjs, 239, 32

\bibitem[{{Pinsonneault} {et~al.}(2025){Pinsonneault}, {Zinn}, {Tayar}, {Serenelli}, {Garc{\'\i}a}, {Mathur}, {Vrard}, {Elsworth}, {Mosser}, {Stello}, {Bell}, {Bugnet}, {Corsaro}, {Gaulme}, {Hekker}, {Hon}, {Huber}, {Kallinger}, {Cao}, {Johnson}, {Liagre}, {Patton}, {Santos}, {Basu}, {Beck}, {Beers}, {Chaplin}, {Cunha}, {Frinchaboy}, {Girardi}, {Godoy-Rivera}, {Holtzman}, {J{\"o}nsson}, {M{\'e}sz{\'a}ros}, {Reyes}, {Rix}, {Shetrone}, {Smith}, {Spoo}, {Stassun}, \& {Wang}}]{APOKASC3}
{Pinsonneault}, M.~H., {Zinn}, J.~C., {Tayar}, J., {et~al.} 2025, \apjs, 276, 69

\bibitem[{{Pr{\v{s}}a} {et~al.}(2016){Pr{\v{s}}a}, {Harmanec}, {Torres}, {Mamajek}, {Asplund}, {Capitaine}, {Christensen-Dalsgaard}, {Depagne}, {Haberreiter}, {Hekker}, {Hilton}, {Kopp}, {Kostov}, {Kurtz}, {Laskar}, {Mason}, {Milone}, {Montgomery}, {Richards}, {Schmutz}, {Schou}, \& {Stewart}}]{IAU}
{Pr{\v{s}}a}, A., {Harmanec}, P., {Torres}, G., {et~al.} 2016, \aj, 152, 41

\bibitem[{{Recio-Blanco} {et~al.}(2024){Recio-Blanco}, {de Laverny}, {Palicio}, {Cassisi}, {Pietrinferni}, {Lagarde}, \& {Navarrete}}]{Recio24}
{Recio-Blanco}, A., {de Laverny}, P., {Palicio}, P.~A., {et~al.} 2024, \aap, 692, A235

\bibitem[{{Recio-Blanco} {et~al.}(2023){Recio-Blanco}, {de Laverny}, {Palicio}, {Kordopatis}, {{\'A}lvarez}, {Schultheis}, {Contursi}, {Zhao}, {Torralba Elipe}, {Ordenovic}, {Manteiga}, {Dafonte}, {Oreshina-Slezak}, {Bijaoui}, {Fr{\'e}mat}, {Seabroke}, {Pailler}, {Spitoni}, {Poggio}, {Creevey}, {Abreu Aramburu}, {Accart}, {Andrae}, {Bailer-Jones}, {Bellas-Velidis}, {Brouillet}, {Brugaletta}, {Burlacu}, {Carballo}, {Casamiquela}, {Chiavassa}, {Cooper}, {Dapergolas}, {Delchambre}, {Dharmawardena}, {Drimmel}, {Edvardsson}, {Fouesneau}, {Garabato}, {Garc{\'\i}a-Lario}, {Garc{\'\i}a-Torres}, {Gavel}, {Gomez}, {Gonz{\'a}lez-Santamar{\'\i}a}, {Hatzidimitriou}, {Heiter}, {Jean-Antoine Piccolo}, {Kontizas}, {Korn}, {Lanzafame}, {Lebreton}, {Le Fustec}, {Licata}, {Lindstr{\o}m}, {Livanou}, {Lobel}, {Lorca}, {Magdaleno Romeo}, {Marocco}, {Marshall}, {Mary}, {Nicolas}, {Pallas-Quintela}, {Panem}, {Pichon}, {Riclet}, {Robin}, {Rybizki}, {Santove{\~n}a}, {Silvelo}, {Smart}, {Sarro}, {Sordo}, {Soubiran}, {S{\"u}veges},
  {Ulla}, {Vallenari}, {Zorec}, {Utrilla}, \& {Bakker}}]{GSPspecDR3}
{Recio-Blanco}, A., {de Laverny}, P., {Palicio}, P.~A., {et~al.} 2023, \aap, 674, A29

\bibitem[{{Rybizki} {et~al.}(2022){Rybizki}, {Green}, {Rix}, {El-Badry}, {Demleitner}, {Zari}, {Udalski}, {Smart}, \& {Gould}}]{Rybizki22}
{Rybizki}, J., {Green}, G.~M., {Rix}, H.-W., {et~al.} 2022, \mnras, 510, 2597

\bibitem[{{Salaris} {et~al.}(1993){Salaris}, {Chieffi}, \& {Straniero}}]{Salaris93}
{Salaris}, M., {Chieffi}, A., \& {Straniero}, O. 1993, \apj, 414, 580

\bibitem[{{Salsi} {et~al.}(2020){Salsi}, {Nardetto}, {Mourard}, {Creevey}, {Huber}, {White}, {Hocd{\'e}}, {Morand}, {Tallon-Bosc}, {Farrington}, {Chelli}, \& {Duvert}}]{Salsi20}
{Salsi}, A., {Nardetto}, N., {Mourard}, D., {et~al.} 2020, \aap, 640, A2

\bibitem[{{Scalo}(1986)}]{Scalo86}
{Scalo}, J.~M. 1986, \fcp, 11, 1

\bibitem[{{Schonhut-Stasik} {et~al.}(2024){Schonhut-Stasik}, {Zinn}, {Stassun}, {Pinsonneault}, {Johnson}, {Warfield}, {Stello}, {Elsworth}, {Garc{\'\i}a}, {Mathur}, {Mosser}, {Hon}, {Tayar}, {Stringfellow}, {Beaton}, {J{\"o}nsson}, \& {Minniti}}]{APOK2}
{Schonhut-Stasik}, J., {Zinn}, J.~C., {Stassun}, K.~G., {et~al.} 2024, \aj, 167, 50

\bibitem[{{Serenelli} {et~al.}(2017){Serenelli}, {Johnson}, {Huber}, {Pinsonneault}, {Ball}, {Tayar}, {Silva Aguirre}, {Basu}, {Troup}, {Hekker}, {Kallinger}, {Stello}, {Davies}, {Lund}, {Mathur}, {Mosser}, {Stassun}, {Chaplin}, {Elsworth}, {Garc{\'\i}a}, {Handberg}, {Holtzman}, {Hearty}, {Garc{\'\i}a-Hern{\'a}ndez}, {Gaulme}, \& {Zamora}}]{APOKASCnaines}
{Serenelli}, A., {Johnson}, J., {Huber}, D., {et~al.} 2017, \apjs, 233, 23

\bibitem[{Stefan(1879)}]{Stefan1879}
Stefan, J. 1879, Sitzungsber. Kaiserl. Akad. Wiss. Math. Naturwiss. Cl. II. Abth., 79, 391

\bibitem[{{Taniguchi} {et~al.}(2026){Taniguchi}, {de Laverny}, {Recio-Blanco}, {Tsujimoto}, \& {Palicio}}]{Twins}
{Taniguchi}, D., {de Laverny}, P., {Recio-Blanco}, A., {Tsujimoto}, T., \& {Palicio}, P.~A. 2026, \aap, 707, A260

\bibitem[{{Taylor}(2005)}]{Topcat}
{Taylor}, M.~B. 2005, in Astronomical Society of the Pacific Conference Series, Vol. 347, Astronomical Data Analysis Software and Systems XIV, ed. P.~{Shopbell}, M.~{Britton}, \& R.~{Ebert}, 29

\bibitem[{{Valle} {et~al.}(2024){Valle}, {Dell'Omodarme}, {Prada Moroni}, \& {Degl'Innocenti}}]{Valle24}
{Valle}, G., {Dell'Omodarme}, M., {Prada Moroni}, P.~G., \& {Degl'Innocenti}, S. 2024, \aap, 690, A327

\bibitem[{{Wenger} {et~al.}(2000){Wenger}, {Ochsenbein}, {Egret}, {Dubois}, {Bonnarel}, {Borde}, {Genova}, {Jasniewicz}, {Lalo{\"e}}, {Lesteven}, \& {Monier}}]{Simbad}
{Wenger}, M., {Ochsenbein}, F., {Egret}, D., {et~al.} 2000, \aaps, 143, 9

\bibitem[{{Zinn} {et~al.}(2022){Zinn}, {Stello}, {Elsworth}, {Garc{\'\i}a}, {Kallinger}, {Mathur}, {Mosser}, {Hon}, {Bugnet}, {Jones}, {Reyes}, {Sharma}, {Sch{\"o}nrich}, {Warfield}, {Luger}, {Vanderburg}, {Kobayashi}, {Pinsonneault}, {Johnson}, {Huber}, {Buder}, {Joyce}, {Bland-Hawthorn}, {Casagrande}, {Lewis}, {Miglio}, {Nordlander}, {Davies}, {De Silva}, {Chaplin}, \& {Silva Aguirre}}]{Zinn22}
{Zinn}, J.~C., {Stello}, D., {Elsworth}, Y., {et~al.} 2022, \apj, 926, 191

\end{thebibliography}

\end{document}